\documentclass[10pt,prd,preprintnumbers,floatfix,aps,notitlepage,showpacs,twocolumn,nofootinbib]{revtex4} 

\usepackage{verbatim}

\usepackage[USenglish]{babel}

\usepackage{amsmath}
\usepackage{amsfonts}
\usepackage{graphicx}

\usepackage{epsfig}
\usepackage{amssymb}

\usepackage[all]{xy} 

\usepackage{hyperref}

\usepackage{latexsym}
\usepackage{epsfig}
\usepackage{amssymb}

\usepackage{fancybox}

\usepackage{wrapfig}
\usepackage{sidecap}

\usepackage{threeparttable}

\newcommand{\hhline}{\noalign{\smallskip} \hline \hline \noalign{\smallskip}}

\newcommand{\lp}{\left(}
\newcommand{\rp}{\right)}
\newcommand{\lb}{\left[}
\newcommand{\rb}{\right]}

\newcommand{\beq}{\begin{eqnarray}}
\newcommand{\eeq}{\end{eqnarray}}
\newcommand{\be}{\begin{eqnarray}}
\newcommand{\ee}{\end{eqnarray}}
\newcommand{\ba}{\begin{eqnarray}}
\newcommand{\ea}{\end{eqnarray}}

\newcommand{\Lag}{\mathcal{L}}

\newcommand{\al}{\alpha}
\newcommand{\bt}{\beta}
\newcommand{\ga}{\gamma}

\newcommand{\la}{\lambda}

\newcommand{\Ga}{\Gamma}

\newcommand{\La}{\Lambda}

\newcommand{\half}{\frac{1}{ 2}}

\newcommand{\ud}[2]{^{#1}_{\phantom{#1} #2}}
\newcommand{\du}[2]{_{#1}^{\phantom{#1} #2}}

\newcommand{\mc}{\mathcal}
\newcommand{\ph}{\phantom{\al}}

\begin{document}

\title{DBI Galileons in the Einstein Frame: Local Gravity and Cosmology}
\author{Miguel Zumalac\'arregui$^{1,2}$}
\author{Tomi S. Koivisto$^2$}
\author{David F. Mota$^2$}
\affiliation{$^{1}$ Instituto de F\'isica Te\'orica IFT-UAM-CSIC, Universidad Aut\'onoma de Madrid, C/ Nicol\'as Cabrera 13-15,
Cantoblanco, 28049 Madrid, Spain}
\affiliation{$^2$ Institute for Theoretical Astrophysics, University of
 Oslo, Sem Sælands vei 13, P.O.\ Box 1029 Blindern, N-0315 Oslo, Norway} %


\begin{abstract}
It is shown that a disformally coupled theory in which the gravitational sector has the Einstein-Hilbert form is equivalent to a quartic DBI Galileon Lagrangian, possessing non-linear higher derivative interactions, and hence allowing for the Vainshtein effect. This Einstein Frame description considerably simplifies the dynamical equations and highlights the role of the different terms.
The study of highly dense, non-relativistic environments within this description unravels the existence of a disformal screening mechanism, 
while the study of static vacuum configurations reveals the existence of a Vainshtein radius, at which the asymptotic solution breaks down.
Disformal couplings to matter also allow the construction of Dark Energy models, which behave differently than conformally coupled ones and introduce new effects on the growth of Large Scale Structure over cosmological scales, on which the scalar force is not screened. We consider a simple Disformally Coupled Dark Matter model in detail, in which standard model particles follow geodesics of the gravitational metric and only Dark Matter is affected by the disformal scalar field. This particular model is not compatible with observations in the linearly perturbed regime. Nonetheless, disformally coupled theories offer enough freedom to construct realistic cosmological scenarios, which can be distinguished from the standard model through characteristic signatures.
\end{abstract}

\maketitle

\section{Introduction}

In the standard $\Lambda$CDM model of cosmology \cite{Komatsu:2010fb}, the universe at the present day appears to be extremely fine tuned. The energy scale of the 
$\Lambda$ component is extremely small compared to the na\"ive quantum corrections \cite{RevModPhys.61.1,Martin:2012bt}, and yet large enough to be detectable through its effect on the cosmological expansion \cite{Perlmutter:1998np,Riess:1998cb,Weinberg:2012es}. This mystery has triggered numerous proposals in which the cosmological constant is exchanged with scalar field sourced dynamical Dark Energy \cite{Copeland:2006wr} or alternative theories of gravity \cite{Clifton:2011jh}.

The set of viable theories is severely limited by Ostrogradski's Theorem \cite{Ostrogradski}. It states that there exists a linear instability in any non-degenerate theory whose fundamental dynamical variable appears in the action with higher than 2nd order in time derivatives: the Hamiltonian for this type of theory is not bounded from below and therefore it accepts configurations with arbitrarily large negative energy \cite{Woodard:2006nt,Chen:2012au}. This result can be bypassed by considering \emph{degenerate} theories, i.e. those in which the highest derivative term can not be written as a function of canonical variables. 
In this case, the dynamics is described by second order equations of motion, even while the action contains higher derivative terms. If gravity only involves a rank two tensor, Lovelock's Theorem \cite{Lovelock:1971yv} states that the Einstein-Hilbert action with a Cosmological Constant is the only theory based on a local,%
\footnote{There is some evidence that there exist viable nonlocal theories ameliorating the fundamental problems of gravity \cite{Biswas:2011ar,Modesto:2011kw}.} 
Lorentz-invariant Lagrangian depending on the metric tensor and its derivatives which gives rise to second order equations of motion in four space-time dimensions.

The addition of a scalar degree of freedom provides a generous extension of the possibilities. The most general gravitational sector for a scalar-tensor theory was first derived by Horndeski \cite{Horndeski} and has received considerable attention recently \cite{Deffayet:2009wt,Deffayet:2009mn,Deffayet:2011gz,Charmousis:2011bf,DeFelice:2011hq,Kobayashi:2011nu,Amendola:2012ky}. It is given by the \emph{Horndeski Lagrangian}
\begin{equation}\label{eq:hornyLagrangian}
 \Lag_H = \sum_{i=2}^{5} \Lag_i\,.
\end{equation}
Up to total derivative terms that do not contribute to the equations of motion, the different pieces can be written as \cite{DeFelice:2011hq}
\begin{eqnarray}
\Lag_2 &=&   G_2(X,\phi)\,, \label{LH2} 
 \\[5pt]
\Lag_3 &=& -G_3(X,\phi) \Box \phi\,, \label{LH3}
 \\[5pt]
\Lag_4 &=& G_4(X,\phi) R + G_{4,X}\lb (\Box\phi)^2 - \phi_{;\mu\nu}\phi^{;\mu\nu} \rb\,, \label{LH4}
\\[5pt]
\Lag_5 &=& G_5(X,\phi) G_{\mu\nu}\phi^{;\mu\nu}  - \frac{1}{6}G_{5,X}\Big[(\Box\phi)^3 \nonumber
\\ && 
- 3(\Box\phi)\phi_{;\mu\nu}\phi^{;\mu\nu} 
+ 2\phi\du{;\mu}{;\nu} \phi\du{;\nu}{;\la} \phi\du{;\la}{;\mu}\Big]\,. \label{LH5}
\end{eqnarray}
Here $R,G_{\mu\nu}$ are the Ricci scalar and the Einstein tensor, $X\equiv -\half g^{\mu\nu}\phi_{,\nu}\phi_{,\mu}$ is the scalar field canonical kinetic term and commas and semi-colon represent partial and covariant derivatives respectively.
On top of a generalized k-essence term (\ref{LH2}), the remaining pieces (\ref{LH3}-\ref{LH5}) fix the tensor contractions, which rely on the anti-symmetric structure of the $\phi_{;\mu\nu}$ terms to trade higher derivatives with the Riemann tensor in the equations of motion. Note that Einstein gravity is recovered by a constant $G_4=M_p^2/2$, while a field dependence $G_4=\omega(\phi)M_p^2/2$ yields an old school scalar-tensor theory, without adding higher derivative interactions (when combined with a suitable kinetic term for the scalar e.g. Brans-Dicke \cite{Brans:1961sx}). 
The theories in which the free functions in (\ref{LH3}-\ref{LH5}) depend on the canonical kinetic term $X$ require the presence of degenerate terms with higher derivatives. Theories for which $G_3,G_4,G_5$ have simple $X-$dependences are usually known as \emph{covariant Galileons} \cite{Deffayet:2009wt,Deffayet:2009mn,Deffayet:2011gz}, while theories with more general $X$ dependence are often known as \emph{generalized Galileons}. Some of the possibilities considered so far are listed in Table \ref{hornytheories}.

In the pursue of generality, one can further consider theories in which the scalar field is allowed to enter the matter sector directly. This type of relation is found in old school scalar-tensor theories, which can be expressed as Einstein's theory, plus a scalar field entering the matter sector by means of a conformal transformation \cite{Clifton:2011jh}. 
Bekenstein studied the most general relation between the physical and the gravitational geometry (i.e. the two metrics out of which the gravitational and the matter Lagrangians are constructed) compatible with general covariance \cite{Bekenstein:1992pj}.
When it only involves a scalar field $\phi$, it is given by the \emph{disformal relation}
\begin{equation} \label{metrics}
\bar{g}_{\mu\nu} = A(\phi)g_{\mu\nu} + B(\phi)\phi_{,\mu}\phi_{,\nu}\,.
\end{equation} 
The free functions $A$ and $B$ may also depend upon the scalar kinetic term $X$ in general, but we will focus on the simpler case here. Previous applications of such a relation to cosmology include varying speed of light theories \cite{Magueijo:2003gj,Bassett:2000wj}, Lorentz invariance violation \cite{Brax:2012hm}, inflation \cite{Kaloper:2003yf}, massive gravity \cite{deRham:2010ik,deRham:2010kj}, Dark Energy \cite{Koivisto:2008ak,Zumalacarregui:2010wj}, relativistic MOND theories \cite{Bekenstein:2004ne,Milgrom:2009gv} and extensions of Dark Matter \cite{Bettoni:2011fs,Bettoni:2012xv}. The present work studies the implications of such a coupling, expanding on our previous analysis \cite{Koivisto:2012za}.

The disformal relation (\ref{metrics}) can also be motivated in theories with extra dimensions, in which matter is confined to a 3+1 dimensional brane embedded in a larger bulk space \cite{deRham:2010eu,Hinterbichler:2010xn,Goon:2010xh,Trodden:2011xh,VanAcoleyen:2011mj,Goon:2011uw,Goon:2011qf,Charmousis:2005ey,Cembranos:2004eb,Cembranos:2003fu,Cembranos:2003mr} (see Ref. \cite{Maartens:2010ar} for a review).%
\footnote{The disformal relation also appears in condensed matter systems, e.g. to study two dimensional lattices such as graphene \cite{PhysRevLett.108.227205}.}
The action for this type of theories is constructed using geometric scalars computed out of the \emph{induced metric}
\begin{equation}\label{branemetric}
 \bar g_{\mu\nu}=g_{\mu\nu} + \pi_{I,\mu}\pi_{,\nu}^I \,,
\end{equation}
where the moduli fields $\pi^I$ represent the coordinates orthogonal to the brane and $g_{\mu\nu}$ is the bulk metric prior to the embedding, necessary to describe gravity.
In the case of a single extra dimension \cite{deRham:2010eu}, the most general Lagrangian contains four terms with a particular form of the Horndeski free functions (\ref{LH2}-\ref{LH5}) and arbitrary prefactors.
The quadratic term is due to the brane tension and has the Dirac-Born-Infeld (DBI) \cite{Alishahiha:2004eh} form, $G_2\propto\sqrt{1+(\partial\pi)^2}$. Therefore, these models are known as DBI Galileons \cite{deRham:2010eu}. The higher order terms arise from curvature invariants computed out to the induced metric (\ref{branemetric}), which produce second order equations of motion \cite{Lovelock:1971yv}: $G_3$ arises from the trace of the extrinsic brane curvature, $G_4$ from the Ricci scalar and $G_5$ from a combination of extrinsic curvature terms and the induced Einstein tensor.
DBI Galileons with more than one extra dimension only accept the generalization of the quadratic and quartic terms $G_2,G_4$ in their Lagrangians \cite{Hinterbichler:2010xn,Charmousis:2005ey}. This restriction is necessary to preserve the symmetry between the directions transverse to the brane - e.g. the moduli fields $\pi_I$ in (\ref{branemetric}).

The usual Galileon terms \cite{Nicolis:2008in} are obtained from DBI Galileon Lagrangian by assuming a flat bulk metric $g_{\mu\nu}\to \eta_{\mu\nu}$ and taking the non-relativistic limit (i.e. low order corrections in $(\partial\pi)^2$). Galileon theories have attracted considerable attention recently \cite{Hui:2010dn,Mota:2010bs,Nesseris:2010pc,Gannouji:2010au,Ali:2010gr,Appleby:2011aa,DeFelice:2011bh,DeFelice:2011aa,RenauxPetel:2011uk,RenauxPetel:2011dv,Brax:2011sv,deRham:2012az,Barreira:2012kk,Hui:2012qt,Hui:2012jb,Appleby:2012ba,deRham:2012fw,Padilla:2012dx,Hiramatsu:2012xj,Ali:2012cv,Sampurnanand:2012wy,Germani:2012qm,Okada:2012mn,Leon:2012mt,Babichev:2012re,Barreira:2013jma} 
because they capture interesting features of higher dimensional models such as DGP \cite{Dvali:2000hr}, including the Vainshtein screening mechanism \cite{Vainshtein:1972sx}. This effect hides the presence of the scalar force due to the non-linear derivative self-interactions of the field, which suppress the field's spatial gradients around matter sources within the so-called Vainshtein radius. The extra force is active on larger distances, potentially having significant cosmological implications.

\begin{table*}
\begin{center}
   \begin{tabular}{ l @{\hspace{20pt}} c @{\hspace{20pt}}  c @{\hspace{20pt}} c @{\hspace{20pt}} c  @{\hspace{20pt}} c  } 
 Theory    & $G_2$ & $G_3$ & $G_4$ & $G_5$  & $g^M_{\mu\nu}$  \\ \hhline
 General Relativity 
      & $\La$ & 0 & $\frac{M_p^2}{2}$ & 0  &  $g_{\mu\nu}$  \\ 
\hhline 
Quintessence 
      & $X+V(\phi)$ & 0 & $\frac{M_p^2}{2}$ & 0  & $g_{\mu\nu}$  \\[5pt]

Generalized k-essence{$^\ddag$} 
      & $K(X,\phi)$ & 0 & $\frac{M_p^2}{2}$ & 0 & $g_{\mu\nu}$  \\ 
\hhline
Old school Scalar-Tensor:
      &  &  & &  &    \\
\, - Jordan Frame 
      & $ X + V(\phi) $ & 0 & $h(\phi)\frac{M_p^2}{2}$ & 0  & $g_{\mu\nu}$  \\[5pt] 
\, - Einstein Frame 
      & $ \tilde X + V(\tilde \phi) $ & 0 & $\frac{M_p^2}{2}$ & 0  & $h^{-1}(\tilde \phi) g_{\mu\nu}$  \\ 
\hhline
 Covariant Galileon{$^\S$}\,  \cite{Deffayet:2009wt}
      & $c_1\phi - c_2 X$ & $\frac{c_3}{M^3}X$ & $ \frac{M_p^2}{2} - \frac{c_4}{M^6}X^2$ & $\frac{3c_5}{M^9}X^2$  & $A(\phi)g_{\mu\nu}$  \\[5pt]
Kinetic Gravity  
Braiding \cite{Deffayet:2010qz,Pujolas:2011he}
     & $K(X,\phi)$ & \hspace{-0.2cm} $G(X,\phi)$ & $\frac{M_p^2}{2}$ & 0  & $g_{\mu\nu}$  \\[5pt]
Purely Kinetic   
Gravity\, \cite{Gubitosi:2011sg}
      & $X$ & 0 & $\frac{M_p^2}{2}$ & $-\la\frac{\phi}{M_p^2}$  &  $g_{\mu\nu}$  \\ \hhline
 DBI Galileon{$^\dag$}\, \cite{deRham:2010eu} 
      & $-\la \gamma^{-1}$ & $ - M_5^3 \ga^2$ & $ \ga^{-1} M_4^2$  &  $-\bt\frac{M_5^2}{m^2}\ga^2$ &  $g_{\mu\nu}$  \\
\hhline
 Disformally Coupled 
 Scalar \cite{Kaloper:2003yf,Koivisto:2012za}   & $X+V(\phi)$ & 0 & 0 & 0  &  $ Ag_{\mu\nu} +  B\phi_{,\mu}\phi_{,\nu} $ 
    \\ 
\hhline
\end{tabular}
\begin{itemize} 
\footnotesize
\item[$^\ddag$] See Table 1 in Ref. \cite{Zumalacarregui:2010wj} for an assortment of k-essence models constructed using disformal relations.
\item[$^\S$] The usual Galileon \cite{Nicolis:2008in} is recovered in the absence of curvature. The analysis of these theories often postulates a conformal coupling between the matter and the field (conformal Galileon).
\item[$^\dag$] References \cite{Goon:2011uw,Goon:2011qf} provide generalizations constructed within the probe brane scheme.
\end{itemize}
\caption[Horndeski projection of Modified Gravity and Dark Energy theories]{Horndeski projection of Modified Gravity and Dark Energy theories. The possibilities shown take into account the Horndeski Lagrangian (\ref{eq:hornyLagrangian}) (see also \cite{DeFelice:2011hq}), the arguments by Bekenstein leading to the disformal metric (\ref{metrics}) and the possibility of defining disformally related frames (see Section \ref{section:frames}). It is then possible to consider a theory of the form $  S_{\rm HB} = \int d^4 x \lp \sqrt{-g} \Lag_H + \sqrt{-g^{M}} \Lag_M(g^{M}_{\mu\nu},\psi)\rp$
as the most general case with a universal coupling to matter.
Here $M_p^2=(8\pi G)^{-1}$, $X=-\frac{1}{2}\phi_{,\mu}\phi^{,\mu}$ and $\gamma=\frac{1}{\sqrt{1-2X}}$ is a brane Lorentz factor. 
\label{hornytheories}}
\end{center}
\end{table*}

This work presents results that may simplify considerably the analysis of theories based on higher dimensional models. Section \ref{section_pointparticle} presents the coupling to point particles and Section \ref{section:frames} introduces theories constructed out of two metrics which are disformally related. In Section \ref{section:confframe} it is argued that by performing a disformal transformation, a theory in which the gravitational sector is standard, but the matter metric is constructed disformally (\ref{metrics}), can be rewritten in a form equivalent to the quartic DBI Galileon Lagrangian (\ref{LH4}), which arises from the scalar curvature computed using the induced metric (\ref{branemetric}) \cite{deRham:2010eu}. Disformally coupled theories therefore provide an Einstein Frame description of certain brane-world constructions, similar to the way in which the field dependent coefficient of the Ricci scalar can be moved from the gravitational to the matter sector in old school scalar-tensor theories. 
In Section \ref{section_equations}, the equations of motion are derived in the Einstein Frame, in which the gravitational sector has the Einstein-Hilbert form but the matter action includes the scalar field as prescribed by Equation (\ref{metrics}). Some properties of the field and the coupling are discussed in Sections \ref{Bp_instab} and \ref{perfectfluid}.

The Einstein Frame description of disformally coupled theories unravels the existence of a \emph{disformal screening mechanism} \cite{Koivisto:2012za}, in which the coupling vanishes if the field is static and the coupled matter behaves as non-relativistic dust. 
Section \ref{section:disfscreening} explores the dynamics in high density, non-relativistic environments for the simpler case of a canonical scalar with a potential and no conformal coupling. In Section \ref{fielddecoupl}, a simple solution is derived in which the field rolls homogeneously regardless of the matter distribution, hence avoiding the formation of spatial gradients that would give rise to an additional force. 
The study of static, spherically symmetric configurations performed in Section \ref{section:vainshtein} reveals the existence a characteristic radius at which the effects of the coupling modify the asymptotic solution. This is analogous to the Vainshtein radius, at which the non-linear derivative self-coupling of the field becomes important, and which lies at the core of the Vainshtein screening mechanism.
Finally, Section \ref{s:signatures} presents some regimes in which the effects of the disformal coupling might be observed.

The equivalence between the disformally coupled theory and a covariant Galileon and the aforementioned results imply that the disformal and Vainshtein screening mechanisms are related. 
These two effects rely on the higher derivative form of the field kinetic terms (Vainshtein) and the kinetic mixing between the field and the coupled matter (disformal). The other available screening mechanisms are essentially different, as they exploit the interplay between the field potential and the coupling to matter:
the chameleon fields rely on the high mass of the field in dense surroundings  \cite{Khoury:2003aq}, and the symmetrons are screened in high ambient density due to their field-dependent coupling \cite{Hinterbichler:2010es}.
Screening mechanisms are central to the construction of alternative theories of gravity in which modifications are allowed to occur over cosmological scales, while the gravitational physics operating in the Solar System are close enough to GR to satisfy current precision tests \cite{Will:2005va}.

The cosmological implications for these models are considered in Section \ref{section:disfcosmo}.
The intensity of the purely disformal coupling is approximately proportional to the scalar field energy density $\rho_\phi$, unlike in the conformally coupled case for which it is proportional to the coupled matter density $\rho_m$.
The equations for non-relativistic coupled fluids are given at both the background and linearly perturbed level, including the analysis of fixed points and an analytic expression for effective gravitational constant on small scales.
The cosmological equations are solved numerically for a simple \emph{Disformally Coupled Dark Matter} model (DCDM), in which Dark Matter is the only coupled species. The model is presented and analyzed in detail in Section \ref{section:examplemodel}, including the computation of Dark Matter and baryonic power spectra.
The simple DCDM model enhances the growth of the coupled Dark Matter density contrast too much to be compatible with observations. However, the model contains significant freedom to provide phenomenologically successful alternatives. Several possibilities to render the model viable are discussed in Section \ref{section:disfviable}. 

We conclude in Section \ref{disfcussion} with a discussion of the main results and future research directions. Appendices \ref{disfappendix}-\ref{dynamicalsystem} contain a summary of disformal relations and some lengthy expressions that were not necessary for the main discussion.
Throughout the present work, quantities computed or constructed using the metric (\ref{metrics}) are denoted ``barred'' or ``disformal''. Quantities constructed using $g_{\mu\nu}$ are denoted as ``unbarred'' and do not involve the scalar field. The metric signature is $(-,+,+,+)$ and units in which the speed of light $c=1$ are assumed unless specified otherwise.

\section{A Test Particle in a Disformal Metric}\label{section_pointparticle}

Let us start with the simple exercise of determining the dynamics of a point-like particle with mass $m$ coupled to the disformal metric (\ref{metrics}).
A Lagrangian density for such a system is given by
\begin{equation} \label{proper_a}
\sqrt{-\bar{g}}\bar{\mathcal{L}}_p = -m \sqrt{-\bar{g}_{\mu\nu}\dot{x}^\mu\dot{x}^\nu}\delta^{(4)}_D(x^{\mu}-x^{\mu}(\lambda))\,,
\end{equation} 
where the dot means derivative with respect to the affine parameter $\la$ along the trajectory $x(\lambda)$ and the correct weight for the delta function has been taken.%
\footnote{The one-dimensional definition of the delta function requires that its generalization to higher dimensions cancels out the tensor density in the integrand $\delta^{(n)}_D(x-x_0) = \frac{1}{\sqrt{-g}}\Pi_a \delta_D(x^\al-x_0^\al)$
(e.g. in spherical coordinates $(r^2\sin\theta)^{-1}\delta^{(3)}(x) = \delta_D(r)\delta_D(\theta)\delta_D(\phi)$).
Hence it does not matter whether $\sqrt{-g}$ or $\sqrt{-\bar g}$ is used in the integration, as long as the delta function is consistent with it.}
%
The effects from the coupling can be seen from the barred four-velocity modulus in (\ref{proper_a})
\begin{equation} \label{proper}
\bar{g}_{\mu\nu}\dot{x}^\mu\dot{x}^\nu = A\dot{x}^2+B(\phi_{,\mu}\,\dot{x^\mu})^2\,.
\end{equation} 
Distances are dilated by the conformal factor $A$, as usual. The disformal factor $B$ gives an additional direction-dependent effect proportional to the projection of the four-velocity along the field gradient.
The equations of motion can be obtained by maximizing the proper time of the particle along its path. The result is the \emph{disformal geodesic equation}
\begin{equation} \label{geodesic}
\ddot{x}^\mu + \bar{\Gamma}^\mu_{\alpha\beta}\dot{x}^\alpha \dot{x}^\beta = 0\,,
\end{equation} 
where the barred Levi-Civita connection has been assumed to be torsion-free and such that the metric compatibility relation holds for barred quantities, i.e. $\bar\nabla_{\al}\bar g_{\mu\nu}=0$.
It can be computed from (\ref{metrics}) and written in terms of unbarred covariant derivatives of the barred metric in a rather compact form
\begin{equation} \label{connection}
\bar{\Gamma}^\mu_{\al\bt}  =  \Ga^\mu_{\alpha\beta} 
+ \bar g^{\mu\la}\lp \nabla_{(\al}\bar g_{\bt)\la} - \half \nabla_\la \bar g_{\al\bt}\rp \,.
\end{equation}
Here the symmetrization is defined as $t_{(\al\bt)}\equiv \frac{1}{2}\lp t_{\al\bt}+t_{\bt\al}\rp$.
No assumption about the dependence of $A,B$ has been made to obtain the above expression, which remains valid if $A,B$ depend on $X$. Note that the difference between the two connections is a tensor, as expected.
Appendix \ref{app:geodesics} shows the expansion of (\ref{connection}) in terms of $A,B$ and its derivatives, which is rather lengthy to be included here. In the case of a constant disformal coupling $B(\phi)=M^{-4}$ with no conformal coupling ($A(\phi)=1$) the equation simplifies considerably:
\begin{equation} \label{barGamma}
 \bar \Gamma^\la_{\mu\nu} = \Gamma^\la_{\mu\nu} 
+  \frac{\phi^{,\la}\phi_{;\mu\nu}}{M^{4}+(\partial\phi)^2} \,.
\end{equation}
Then in the non-relativistic limit $\dot x^{i} \sim v/c \ll 1 \approx \dot x^0$, the force produced by such a coupling is $\vec F \propto \ddot\phi \vec\nabla \phi /M^{4}$. This is essentially different from the fifth force produced by a conformally coupled field $\vec F \propto (\log A)_{,\phi}\vec\nabla\phi $.

The stress energy tensor with respect to the unbarred metric can be computed by variation of (\ref{proper_a}) with respect to $g_{\mu\nu}$
\begin{equation} \label{Tdisfpart}
T_p^{\mu\nu} \equiv  \frac{2}{\sqrt{-g}}\frac{\delta\lp\sqrt{-\bar{g}}\,\bar{\mathcal{L}}_{m}\rp}{\delta g_{\mu\nu}}
= A m \frac{\dot{x}^\mu\dot{x}^\nu}{\sqrt{g\dot{\bar{x}}^2}} \delta^{(4)}_D\lp x^{\mu}-x^{\mu}(\lambda)\rp\,.
\end{equation}
If the gravitational metric is the unbarred one, this is the energy momentum tensor sourcing the space-time geometry.%
\footnote{It is possible to write (\ref{Tdisfpart}) in the perfect fluid form $T^{\mu\nu}\equiv \rho u^\mu u^\nu$
if the coupled matter four velocity and the energy density are identified with $u_\mu=\dot{x}_\mu/\sqrt{-\dot{x}^2}$ and
$\rho =  m  \delta^{(4)}_D\lp x^{\mu}-x^{\mu}(\lambda)\rp \sqrt{ \frac{\dot{x}^2}{g}} A^{1/2}
\lp 1- \frac{B}{A}(u^\mu\phi_{,\mu})^2 \rp^{-1/2}$.} 
This result can be used to express the particle Lagrangian in terms of the energy momentum tensor
\begin{equation} \label{lp}
\sqrt{-\bar g} \mathcal{ \bar L}_p  = T_p + \frac{B}{A}\phi_{,\mu}\phi_{,\nu}T^{\mu\nu}_p = \bar g_{\mu\nu}T^{\mu\nu}_p \,.
\end{equation} 
The above expression gives an effective form for the coupling to matter. It shows how the kinetic term of the scalar mixes with the matter content, a very important property that lies at the heart of disformally coupled theories, including the disformal screening mechanism explored in Section \ref{section:disfscreening}.

\section{Disformally Related Theories} \label{section:frames}

\begin{table*}
\begin{displaymath}
 \xymatrix{ 
 Ag_{\mu\nu} \subset \sqrt{-g}\Lag_M   
\ar@{<.>}[r] \ar@{<.>}[d]
    &  \doublebox{  \rm Einstein \; (\ref{einsteinframe})} 
\ar[dr]|{\;\; g_{\mu\nu}\to A^{-1}g_{\mu\nu}} 
\ar[dl]|{g_{\mu\nu}\to g_{\mu\nu} - \frac{B}{A}\phi_{,\mu}\phi_{,\nu}\;\;} 
\ar[dd]|{\textstyle g_{\mu\nu}\to \frac{1}{A}g_{\mu\nu} - \frac{B}{A}\phi_{,\mu}\phi_{,\nu}}
& 
B\phi_{,\mu}\phi_{,\nu} \subset \sqrt{-g}\Lag_M   
\ar@{<.>}[l] \ar@{<.>}[d]
\\
    \boxed{\rm Galileon \; (\ref{confframe})} 
\ar[dr]|{g_{\mu\nu}\to A^{-1}g_{\mu\nu}}
& \hspace{5cm}
& \boxed{\rm Disformal \; (\ref{disfframe})}  
\ar[dl]|{g_{\mu\nu}\to g_{\mu\nu}-B\phi_{,\mu}\phi_{,\nu}}
\\
B\phi_{,\mu}\phi_{,\nu} \subset \sqrt{-g}R   
\ar@{<.>}[r] \ar@{<.>}[u]
    &	\doublebox{\rm Jordan \; (\ref{jordanframe})} & 
Ag_{\mu\nu} \subset \sqrt{-g}R 
\ar@{<.>}[l] \ar@{<.>}[u]
%
  } 
\end{displaymath}

\caption[Physical frames for disformally coupled theories]{\justifying %
Physical frames for disformally coupled theories. The intermediate frames are named after the effects of the disformal coupling, e.g. a disformal coupling in the matter sector (Disformal Frame) and a quartic Galileon term in the gravity sector (Galileon Frame). The transformation rules (solid arrows) are based on the action (\ref{mastertheory}) and given in terms of the definitions (\ref{einsteinframe}-\ref{jordanframe}). Note that the transformations commute. Dotted arrows indicate whether the conformal and disformal parts of the coupling enter the gravitational or the matter sector for each given frame.} \label{disformalframes}
\end{table*}

The previous Section presented a simple example of a theory in which the matter Lagrangian is constructed using a disformal metric (\ref{metrics}). Although no gravitational sector was specified, the simplest possibility is to assume that it is given by the Einstein-Hilbert form computed out of the unbarred metric $g_{\mu\nu}$. In this case, Einstein equations retain the usual form and are sourced by the energy momentum tensor (\ref{Tdisfpart}). We shall refer to disformally coupled theories in which the gravitational sector is standard as being expressed in the \emph{Einstein Frame} (EF), in analogy with old school scalar-tensor theories.
More generally, one wishes to know what kind of theories can be constructed using two metrics that are disformally related and study the connections between them. This generalizes the conformal equivalence between old school scalar-tensor theories minimally coupled to matter and theories with a standard gravitational sector, but with a non-minimal coupling between matter and the scalar.

In order to consider theories which allow an Einstein Frame description, one starts with a general \emph{bi-metric theory} where the gravity sector has the EH form, but with unspecified forms for the gravitational and matter metrics 
\begin{equation}\label{mastertheory}
 S = \int d^4 x \lp \sqrt{-g^{G}}R\lb g^{G}_{\mu\nu}\rb - \sqrt{-g^{M}}\Lag_m\lp g^{M}_{\mu\nu},\psi\rp \rp \,.
\end{equation}
Playing with the disformal relations between $g^{G}_{\mu\nu}$ and $g^{M}_{\mu\nu}$ allows one to write the above theory in different frames. Besides the Einstein Frame, an obvious possibility is to consider the \emph{Jordan Frame} (JF), a description in which matter appears minimally coupled and the field only enters the gravitational sector. But since the disformal coupling has two parts, two more \emph{intermediate frames} can be defined, in which only a certain part of the coupling enters the matter action. 
The four possibilities are described below and summarized in Table \ref{disformalframes}, together with the transformations that provide the connections between them.
For the sake of simplicity, the Einstein Frame has been defined using a matter metric of the form (\ref{metrics}), consistently with the notation used in most of the paper.

\begin{enumerate}
 \item {\bf Einstein Frame:}
\begin{equation}
\label{einsteinframe}
 g^{G}_{\mu\nu}= g_{\mu\nu},\quad\quad g^{M}_{\mu\nu} = A g_{\mu\nu} + B\phi_{,\mu}\phi_{,\nu}\,.
\end{equation}
This is the formulation used throughout the rest of the paper. The equations in this frame are derived in Section \ref{section_pointparticle} for a point particle and in Section \ref{section_equations} in general.

\item {\bf Disformal Frame:} 
\begin{equation}\label{disfframe}
 g^{G}_{\mu\nu}= \frac{1}{A} g_{\mu\nu},\quad\quad g^{M}_{\mu\nu} = g_{\mu\nu} + B\phi_{,\mu}\phi_{,\nu}\,.
\end{equation}
The disformal part enters the matter Lagrangian explicitly. The conformal factor enters the gravitational sector through a coupling to $R$, like in old school scalar-tensor theories.

\item {\bf Galileon Frame:} 
\begin{equation}\label{confframe}
 g^{G}_{\mu\nu}= g_{\mu\nu} - \frac{B}{A}\phi_{,\mu}\phi_{,\nu},\quad\quad g^{M}_{\mu\nu} = A g_{\mu\nu}\,.
\end{equation}
The conformal part enters the matter Lagrangian explicitly and the field couples directly to gravity as a DBI Galileion, see Section \ref{section:confframe}.

\item {\bf Jordan Frame:} 
\begin{equation}\label{jordanframe}
 g^{G}_{\mu\nu}= \frac{1}{A}g_{\mu\nu} - \frac{B}{A}\phi_{,\mu}\phi_{,\nu},\quad\quad g^{M}_{\mu\nu} = g_{\mu\nu}\,.
\end{equation}
Matter is minimally coupled to a metric and the field enters the gravitational sector exclusively. 
\end{enumerate}
The JF is the most convenient frame to analyze certain properties of the theory and its predictions, as matter follows the geodesics of the simple metric $g_{\mu\nu}$.
The matter metric in the remaining frames contains the scalar field explicitly, and therefore matter moves along geodesics that involve the field variations (\ref{geodesic}) in these representations. These frames are still interesting to analyze the theory. For example, the equations simplify considerably in the EF, just like in conformally related theories. Once these are solved, the solutions can be used to write down the Jordan Frame metric. 

The explicit computation of the curvature scalar for a metric which includes a disformal part allows one to connect the theory studied in the Einstein Frame with a particular sector of the Horndeski Lagrangian (\ref{eq:hornyLagrangian}). As anticipated in the introduction, the thus obtained theory is related to a type of DBI covariant Galileon when expressed in the Galileon or Jordan Frames.

\subsection{Disformal Curvature: The Galileon Frame} \label{section:confframe}

It is possible to get a sense of disformally coupled theories in a different frame by applying the transformations sketched in Table \ref{disformalframes} to known actions.%
\footnote{The authors of Ref. \cite{Charmousis:2012dw} define a Galileon Frame through a conformal transformation, which is therefore essentially different from the one considered here.}
The simplest case is the canonical scalar field as was described in Ref. \cite{Zumalacarregui:2010wj}, where it was shown that the transformation produced the disformal quintessence Lagrangian
\begin{eqnarray} \label{disfquint}
 && \sqrt{-g}(X - V) \xrightarrow{g_{\mu\nu}\to\bar g_{\mu\nu}} \sqrt{-\bar g}( \bar X - V)=  \\
 && \phantom{\sqrt{-g}(X } = A^{3/2}\sqrt{-g}\lp \frac{X}{\sqrt{A-2BX}} - \sqrt{A-2BX} V\rp
\nonumber \,,
\end{eqnarray}
up to the ambiguity in the definition of the kinetic term described in Section 2.4 of Ref. \cite{Zumalacarregui:2010wj}. The above theory encompasses several Dark Energy models in certain limits, which are obtained by appropriate choices of $A,B$ and $V$.

Considering similar relations when the disformal transformations involve the gravitational sector in (\ref{mastertheory}) requires the computation of the Ricci curvature for a barred metric that includes the scalar field as in (\ref{metrics}). The starting point is the difference between the standard and the barred connection (\ref{connection})
\begin{eqnarray} \label{connections}
\mathcal{K}^{\al}_{\ph \mu\nu} &\equiv&  \bar\Gamma^{\al}_{\mu\nu} - \Gamma^{\al}_{\mu\nu}
= \bar g^{\al\la} \Big( \nabla_{(\mu}\bar g_{\nu)\la} -\half\nabla_\la \bar g_{\mu\nu} \Big)
\end{eqnarray}
where the symmetrization is defined as $t_{(\al\bt)}\equiv \frac{1}{2}\lp t_{\al\bt}+t_{\bt\al}\rp$. 
The barred Riemann tensor is obtained from the usual definition, 
and it can be related to the unbarred one in a manifestly tensorial form in terms of (\ref{connections})
\begin{eqnarray}\label{riemmangen}
 \bar R^\al_{\phantom{\al}\bt\mu\nu} 
&\equiv& \partial_{[\mu}\bar\Gamma^\al_{\nu]\bt} + \bar\Gamma^\al_{\ga[\mu}\bar\Gamma^\ga_{\nu]\bt} \label{riemann}\\ 
&=&  R^\al_{\ph\bt\mu\nu}
 + \nabla_{[\mu}\mc K^{\al}_{\ph \nu]\bt}
+ \mc K ^\al_{\ph \gamma [\mu} \mc K^\gamma _{\ph \nu]\bt}\,,  \nonumber 
\end{eqnarray}
where anti-symmetrization is defined without the usual $\frac{1}{2}$ coefficient $A_{[\al\bt]}\equiv A_{\al\bt}-A_{\bt\al}$. The Ricci scalar follows from the contraction 
\begin{equation}\label{riccigen}
\bar R \equiv \bar g^{\mu\nu} \bar R^\al_{\ph\mu\al\nu}  \,.
\end{equation}
with the inverse barred metric 
\begin{equation}\label{inverse}
 g^{\mu\nu}= \frac{1}{A}\lp g^{\mu\nu} - \frac{B}{A-2BX}\phi^{,\mu}\phi^{,\nu}\rp\,.
\end{equation}
Finally, the disformal Einstein-Hilbert Lagrangian density requires the barred volume element (\ref{determinant2}) to be covariant
\begin{equation}\label{determinant}
\sqrt{-\bar{g}}= \sqrt{-g} A^2\sqrt{1-2\frac{B}{A}X} \,.
\end{equation} 
Note that no assumption has been made about the functions $A,B$ out of which the geometric quantities (\ref{connections}-\ref{riccigen}) are computed. However, the general computation is very lengthy, and it is useful to adopt some simplifications.

Let us focus for the time being on a theory in the Galileon Frame, for which the disformal part is absorbed into the gravitational sector.
The following computation assumes thus a gravitational metric of the form (\ref{confframe})
\begin{equation} \label{metrics2}
g^G_{\mu\nu} \equiv \bar g_{\mu\nu} = g_{\mu\nu} + D(\phi)\phi_{,\mu}\phi_{,\nu}\,, 
\end{equation}
where the disformal factor $D(\phi)$ is yet to be specified.
\footnote{Just as in the rest of the paper, $D$ has been assumed to be independent of the field derivatives $X$. This assumption is important in order to simplify the computations performed below. However, given the importance of the $X$ dependence of higher derivative terms in the Horndeski Lagrangian (\ref{eq:hornyLagrangian}) and related theories, it is worth considering the more general case $D(\phi,X)$, although in this case the equations would become very involved.}
The Jordan Frame can be obtained at the end of the computation by inverting the conformal transformation $g_{\mu\nu}\to A^{-1}(\phi)g_{\mu\nu}$ in the resulting metric and curvature objects. Since the transformation rules for curvature tensors under conformal relations are well known \cite{Carroll:2004st}, and Galileon-like theories usually retain a conformal coupling to matter in phenomenological applications (e.g. \cite{Andrews:2010km}), the Jordan Frame curvature will not be computed explicitly.

The barred metric (\ref{metrics2}) can be simplified by a redefinition of the field
\begin{eqnarray}\label{disfrescale}
 \bar g_{\mu\nu}=g_{\mu\nu} + \pi_{,\mu}\pi_{,\nu}\,, \; \text{ with } \pi \equiv \int\sqrt{D(\phi)}d\phi \,,
\end{eqnarray}
where we have assumed that $B(\phi) \geq 0$.
The above expression has the same form as the effective metric in probe-brane theories (\ref{branemetric}). It simplifies considerably the computation of the connection tensor (\ref{connections}), which now reads
\begin{eqnarray}
\mathcal{K}^{\al}_{\ph \mu\nu} = \bar g^{\al\la}\lp \pi_{,\la}\pi_{;\mu\nu}\rp = \ga^2 \pi^{,\al}\pi_{;\mu\nu}  \,. \label{simpleconn}
\end{eqnarray}
Here
\begin{equation}
\ga=\frac{1}{\sqrt{1+\pi_{,\al}\pi^{,\al}}}\,,
\end{equation}
is a Lorentz factor that arises from the inverse metric (\ref{inverse}).
The barred Riemann tensor can be computed directly from (\ref{riemann}), rewriting anti-symmetrized derivatives in terms of the curvature $\nabla_{[\mu}\nabla_{\nu]}X^{\cdots\al_i\cdots}_{\cdots\bt_j\cdots} = \Sigma_i R^{\al_i}_{\ph\la\mu\nu}X^{\cdots\la\cdots}_{\cdots\cdots}
-  \Sigma_j R^{\la}_{\ph\bt_j \mu\nu}X_{\cdots\la\cdots}^{\cdots\cdots}$. 
The result is also simple
\begin{equation}
\bar R^\al_{\ph\bt\mu\nu} = \bar g^{\al\la}\lp R_{\la\bt\mu\nu} + \ga^2 \pi_{;\la[\mu}\pi_{;\nu]\bt} \rp  \,.
\end{equation}
The barred Ricci scalar can be easily obtained by a second contraction
\begin{equation}\label{barR}
\bar R = \lp g_{\mu\nu} - 2\ga^2 \pi_{,\mu}\pi_{,\nu} \rp
\Big[ R^{\mu\nu} + \ga^2 \lp \pi^{;\mu\nu}\Box\pi - \pi^{;\mu}_{\ph ;\al}\pi^{;\al\nu}\rp \Big]\,.
\end{equation}
Finally, the gravitational Lagrangian $\sqrt{-\bar g}\bar R$ just requires multiplying by the barred volume factor (\ref{determinant}) to make it covariant.

The total action for the theory in the Galileon Frame is obtained by adding a matter Lagrangian with a conformal factor $A(\phi)$ in the matter metric 
\begin{equation}
S_{\rm GF}= \int d^4 x \sqrt{-g}\left\{ \frac{M_p^2}{2} \Lag_{\rm GF}
+ A^2\Lag_m(A g_{\mu\nu},\psi) \right\}  \,, 
\end{equation}
where  $\Lag_{\rm GF} \equiv \sqrt{\frac{\bar g}{g}} \bar R$ reads
\begin{eqnarray}
\Lag_{\rm GF}  &=& 
\frac{1}{\ga}R - 2\ga \pi_{,\mu}\pi_{,\nu}R^{\mu\nu}  
+ \ga \big( (\Box\pi)^2 - \pi_{;\mu\nu}\pi^{;\mu\nu}\big)
\nonumber \\ &&  
-2 \ga^3 \big( \pi_{,\mu}\pi^{;\mu\nu}\pi_{,\nu} \Box\pi - \pi^{,\mu}\pi_{,\mu\al}\pi^{;\al\nu}\pi_{,\nu} \big)
\,. \label{LagGF} 
\end{eqnarray}
These results were previously obtained by de Rham and Tolley \cite{deRham:2010eu} in the context of higher dimensional gravity theories. In particular, the above action corresponds to a quartic DBI Galileon term. This expression can be rewritten in a much simpler form up to a total derivative:%
\footnote{This can be done by substracting the divergence of $\xi^\al= 2\ga (\pi^{\al}\Box\pi - \pi^{;\al\bt}\pi_{,\al})$ and using the fact that $\nabla_{\mu}\ga=-\ga^3 \pi^{,\al}\pi_{;\al\mu}$ and $\pi^{,\bt}\nabla_{[\al}\nabla_{\bt]}\pi^{,\al}= R_{\al\bt}\pi^{,\al}\pi^{,\bt}$.}
\begin{equation}
\Lag_{\rm GF} = 
\frac{1}{\ga}R - \ga \big( (\Box\pi)^2 - \pi_{;\mu\nu}\pi^{;\mu\nu}\big) 
 \,. \label{LagGF2} 
\end{equation}
It is now clear that the action (\ref{LagGF2}) has the right form of the Horndeski Lagrangian (\ref{LH4}), with $G_4=\ga^{-1}=\sqrt{1-2X}$. It reduces to the quartic covariant Galileon term in the non relativistic limit limit $X\equiv -\half\pi_{,\mu}\pi^{,\mu}\ll1$, with the right non-minimal coupling to gravity to yield second order equations of motion.

Actions (\ref{LagGF}, \ref{LagGF2}) are formed by one of the possible terms yielding second order equations of motion. The reader is referred to Section 5 of Ref. \cite{deRham:2010eu} for results on other curvature invariants leading to the cubic and quintic DBI Galileon terms for probe branes with a single extra dimension (See also Section 3.4 of \cite{deRham:2012az} for a summary). If the number of extra dimensions is larger than one, only (\ref{LagGF2}) and the quadratic DBI brane tension terms $G_2 = -\la\ga^{-1}$ are allowed, as they respect the symmetry between the directions transverse to the brane \cite{Hinterbichler:2010xn,Charmousis:2005ey}.

%


The disformal transformation of the metric is analogous to the one used in the analysis of (perturbatively) ghost-free massive gravity proposed by De Rham \emph{et al.} \cite{deRham:2010kj,deRham:2010ik}. In the decoupling limit, the helicity-0 mode is described by a scalar field with non-linear derivative self-interactions. It is possible to write the interactions for the helicity-0 mode as a Galileon Lagrangian by means of a disformal transformation of the metric (e.g. Eq. (35) of Ref. \cite{deRham:2012az}), therefore expressing the theory in a frame which features a conformal as well as a disformal coupling to matter. However, massive gravity is more general than this theory, as it contains other interactions between the different degrees of freedom which are not encoded in the Galileon Lagrangian.

The Jordan Frame expression of covariant Galileons was considered by Appleby \& Linder \cite{Appleby:2011aa}, where both the conformal and the disformal coupling to matter were shifted to the gravitational side of the action. The theory studied there is essentially different from (\ref{LagGF2}), as it includes all the standard Galileon terms (which are obtained from the DBI Galileon terms in the limit of small $X$), instead of considering only the quartic term (\ref{LagGF}). 
In the context of inflation, Renaux-Petel \emph{et al.} \cite{RenauxPetel:2011uk,RenauxPetel:2011dv} studied a theory including quartic DBI Galileon terms in the Jordan Frame. Their model also included an Einstein-Hilbert term computed out of the unbarred metric. Therefore, it does not allow for the construction of an Einstein Frame and essentially differs from the theory considered here. However, their study stresses the bi-metric structure of such theories and uses interesting techniques to analyze the dynamical equations.

\section{Equations in the Einstein Frame}\label{section_equations}

In this Section, the equations for a disformally coupled theory which admits an Einstein Frame description will be derived. Such a theory is given by the following action
\begin{equation} \label{disfcoup:action}
S_{\rm EF} = \int d^4x \lb \sqrt{-g}\lp \frac{R[g_{\mu\nu}]}{16\pi G} + \Lag_{\phi} \rp
+ \sqrt{-\bar{g}}\bar{\mathcal{L}}_{m}(\bar g_{\mu\nu},\psi) 
\rb \,.
\end{equation}
The interacting matter sector $\sqrt{-\bar g}\bar \Lag_{m}$ is to be constructed using the barred metric (\ref{metrics}), as it has been made explicit in the second line using the explicit form of $\bar g_{\mu\nu}$ and its determinant (\ref{determinant})
\begin{eqnarray}
&&\sqrt{-\bar{g}}\bar{\mathcal{L}}_{m}(\bar g_{\mu\nu},\psi)  =  
 \\
&& \phantom{\sqrt{-\bar{g}}} = \sqrt{-g} A^2\sqrt{1-2\frac{B}{A}X}  \bar\Lag_m \big( Ag_{\mu\nu} + B\phi_{,\mu}\phi_{,\nu}, \psi \big) \,. \nonumber
\end{eqnarray}
It will be further assumed that there is no dependence on the barred metric derivatives.%
\footnote{This implies that the disformal connection (\ref{connection}) does not appear in the action. This assumption holds for scalar fields and gauge vectors vanishes due to the lack of indices and antisymmetry of the kinetic term, respectively. Although other fields may couple to $\bar g_{\mu\nu,\la}$, the assumption simplifies the equations considerably and is common in the analysis of scalar-tensor theories.}
A scalar field Lagrangian density of the k-essence type $\Lag_{\phi}=\Lag_{\phi}(\phi,X)$ has been also included. More general dependence on the field derivatives may be considered, but this term gives relatively simple equations of motion, and is general enough to accommodate both a canonical $\Lag_\phi = X-V$ and a disformally self-interacting scalar field \cite{Koivisto:2008ak,Zumalacarregui:2010wj}. The matter Lagrangian may include other pieces with different couplings. An uncoupled matter sector can be included by the addition of a Lagrangian $\sqrt{-{g}}{\mathcal{L}}_{u}$ constructed out of the unbarred metric.

The stress energy tensors for both species will be further defined in terms of the contravariant gravitational metric
\begin{eqnarray}
T^{\mu\nu}_{\phi} & \equiv & \frac{2}{\sqrt{-g}}\frac{\delta\lp\sqrt{-g}\,\Lag_{\phi}\rp}{\delta g_{\mu\nu}}\,, \label{Tdisfphi} \\
T^{\mu\nu}_{m} &\equiv & \frac{2}{\sqrt{-g}}\frac{\delta\lp\sqrt{-\bar{g}}\,\bar{\mathcal{L}}_{m}\rp}{\delta g_{\mu\nu}} \,, \label{Tdisfmatt}
\end{eqnarray}
such that the Einstein field equations take the usual form $G^{\mu\nu}=8\pi G T^{\mu\nu}$ and the \emph{total} energy-momentum is covariantly conserved with respect to the unbarred metric by virtue of the Bianchi identities: $\nabla_\mu(T^{\mu\nu}_{m}+T^{\mu\nu}_\phi ) = 0$.
However, the coupling causes that this relation does not occur for each component separately, and in general 
\begin{equation}
\nabla_\mu T^\mu_{(\phi)\nu} =  -\nabla_\mu T^\mu_{(m)\nu} = Q\phi_{,\nu}\,,
\end{equation}
where the form of the interaction (last equality) can be seen by explicitly computing the divergence of the scalar field stress tensor:
\begin{equation}
\nabla_\mu T^\mu_{(\phi)\nu} = 
\lp \Lag_{\phi,\phi}-\nabla_\mu \frac{\partial \Lag_{\phi}}{\partial \phi_{,\mu}}\rp\phi_{,\nu} \equiv Q\phi_{,\nu} \,.
\end{equation}
As the term in parenthesis is equal to the Lagrangian field variation ${\delta\Lag_\phi}/{\delta\phi}$, the equation for the scalar field, ${\delta S}/{\delta \phi} = {\delta \Lag_{\phi}}/{\delta\phi} + {\delta \Lag_{m}}/{\delta\phi}=0$,
allows one to write $Q$ in terms of the variation of the matter Lagrangian
\begin{equation} \label{divergence}
 Q = \frac{1}{\sqrt{-g}}\lp 
\nabla_\mu \frac{\partial \lp \sqrt{-\bar{g}} \mathcal{\bar L}_{m}\rp}{\partial \phi_{,\mu}}
- \sqrt{-\bar{g}} \mathcal{\bar L}_{m,\phi} \rp \,.
\end{equation}
The coupling can be evaluated by application of the chain rule. For the specific form of the barred metric (\ref{metrics}), the variation with respect to the field yields
\begin{equation}\label{phivar}
\frac{\delta \lp\sqrt{-\bar{g}} \mathcal{\bar L}_{m}\rp}{\delta g_{\mu\nu}}\frac{\partial g_{\mu\nu}}{\partial \bar{g}_{\mu\nu}}
\frac{\partial \bar{g}_{\mu\nu}}{\partial \phi} = \frac{\sqrt{-g}}{2A} T^{\mu\nu}_{m} \lp A' g_{\mu\nu}+B' b_\mu b_\nu\rp \,,
\end{equation}
and similarly for the derivative with respect to the field gradient
\begin{equation}\label{phidotvar}
\frac{\delta \lp\sqrt{-\bar{g}} \mathcal{\bar L}_{m}\rp}{\delta g_{\mu\nu}}\frac{\partial g_{\mu\nu}}{\partial \bar{g}_{\mu\nu}}
\frac{\partial \bar{g}_{\mu\nu}}{\partial \phi_{,\mu}} = \sqrt{-g}\frac{B}{A} T^{\mu\nu}_{m} \phi_{,\nu} \,,
\end{equation}

The coupling can be obtained after replacing (\ref{phivar},\ref{phidotvar}) in (\ref{divergence}),
\begin{equation}
 \label{p_div}
Q= \nabla_\mu\lp\frac{B}{A} T^{\mu\nu}_{m} \phi_{,\nu} \rp  
-\lb \frac{A'}{2A}T_{m} + \frac{B'}{2A} \phi_{,\mu} \phi_{,\nu} T^{\mu\nu}_{m} \rb \,.
\end{equation}
The equations for the coupled matter component and the field are then
\begin{eqnarray}
&& \nabla_\mu T^{\mu\nu}_{m} = -Q\phi^{,\nu}   \label{mattereq} \,, \\
&& \mathcal M_{(\phi)}^{\mu\nu} \phi_{;\mu\nu} 
  +  \Lag_{\phi,\phi}- 2X \Lag_{\phi,X\phi} = Q \label{fieldeq1} \,,
\end{eqnarray}
where $\mathcal M_{(\phi)}^{\mu\nu}\equiv \lp \Lag_{\phi,X} g^{\mu\nu} +\Lag_{\phi,XX}\phi^{,\mu}\phi^{,\nu}\rp$ is the general kinetic term for the scalar.
Einstein equations $G^{\mu\nu}=8\pi G (T_{m}^{\mu\nu}+T_{(\phi)}^{\mu\nu})$ together with (\ref{mattereq}, \ref{fieldeq1})
and (\ref{p_div}) determine unambiguously the evolution of matter, the scalar field and the metric. These equations naturally contain the case of a conformally coupled field, where only the coupling to the trace of energy momentum is present in $Q$. Note that so far this result is general and does not depend upon the matter content as long as the matter action only depends on the field through the barred metric (\ref{metrics}) algebraically.


\subsection{Properties of the Scalar Field Equation} \label{Bp_instab} 

The first term in the coupling (\ref{p_div}) contains higher derivatives of the variables $T\ud{\mu\nu}{;\alpha},\phi_{;\mu\nu}$ due to the kinetic mixing in the mater action, cf. (\ref{lp}). These have to be solved for in order to integrate the evolution equations (\ref{mattereq}, \ref{fieldeq1}), which can be done after adopting a coordinate system. It is possible to eliminate the matter derivatives in the scalar field equation by contracting (\ref{mattereq}) with $\phi^{,\nu}$ and solving for $\phi_{,\nu}\nabla_\mu T_{m}^{\mu\nu}$. 
The result can be inserted back in (\ref{fieldeq1}) and rearranged as
\begin{equation}\label{cov-field}
\mathcal M^{\mu\nu}\nabla_\mu\nabla_\nu\phi + \frac{A}{A-2BX}\mathcal Q_{\mu\nu}T^{\mu\nu}_{m} + \mathcal V=0 \,,  
\end{equation}
were we have defined:
\begin{eqnarray}
& \mathcal M^{\mu\nu} &\equiv \Lag_{\phi,X} g^{\mu\nu} + \Lag_{\phi,XX}\phi^{,\mu}\phi^{,\nu} - \frac{B\, T^{\mu\nu}_{m}}{A-2BX}\,, \label{cov-field-mass}  \\
& \mathcal Q_{\mu\nu} &\equiv \frac{A'}{2A}g_{\mu\nu} + \left(\frac{A'B}{A^2} - \frac{B'}{2A}\right) \phi_{,\mu}\phi_{,\nu} \,, \label{Qpot} \\
& \mathcal V &\equiv \Lag_{,\phi} -2X \Lag_{,X\phi} \,. \label{phipot}
\end{eqnarray}
This equation can then be used instead of (\ref{fieldeq1}) to determine the evolution of the scalar field. It displays very clearly the role of the coupling, which enters not only as a modification to the effective potential (second term), but also in the coefficient for the higher derivatives of the field. This feature will be ultimately responsible for the screening mechanism that these models exhibit in high density regions, which is explored in Section \ref{section:disfscreening}.

Equation (\ref{cov-field}) also shows that the different components of the energy momentum tensor may modify the signs of the coefficients of the second order field derivatives. It is then necessary to analyze whether the scalar field propagation has a good initial value formulation, as was pointed out and first analyzed by Bruneton \& Esposito-Far\`ese \cite{Bruneton:2007si}.%
\footnote{Disformally coupled theories are discussed in Section IIIC of ref. \cite{Bruneton:2007si} from a field theoretical perspective. There it is argued that some dependence on $X$ on either the conformal or the disformal factor is necessary for the theory to have a Lorentzian signature for all values of $X$ (cf. their eq. 3.22). However, the dynamics of the scalar field might prevent it from acquiring arbitrarily values of $X$, as it is argued below and shown explicitly for a cosmological model in which $2BX <1$ at any time (See Figure 1). Ref. \cite{Bruneton:2007si} also studies the hyperbolic condition within dynamical pressureless matter, which is complementary to the analysis presented below.}
Equation (\ref{cov-field}) is a quasi-linear, diagonal second order equation \cite{Wald:1984rg}, of the form 
\begin{equation}
 \phi_{;\mu\nu}\mathcal M^{\mu\nu}(\phi,\phi_{,\la},T^{\al\bt}) + f(\phi,\phi_{,\la},T^{\al\bt})=0\,.
\end{equation}
However, its hyperbolic character relies on the signature of $\mc M^{\mu\nu}$, which involves the coupled matter energy-momentum tensor. For a canonical scalar field  $\Lag_\phi = X-V$ disformally coupled to a perfect fluid, the derivatives tensor reads $\mathcal M\ud{\mu}{\nu} =\delta^\mu_\nu -\frac{B}{A-2BX}\mathop{\mathrm{diag}}(-\rho,p,p,p)$ in coordinates comoving with the fluid. 
Positive energy density keeps the correct sign of the time derivative term if $B>0$, avoiding the existence of ghosts modes.
However, a large pressure can flip the sign of the spatial derivatives coefficient, introducing a gradient instability. This might have important consequences in sufficiently relativistic environments.

Addressing the viability of the theory hence requires determining under which conditions the instability may occur dynamically, which in turn requires considering the evolution of the coupled matter components including the non-linear terms in (\ref{p_div}, \ref{cov-field}). 
In certain cases, the system might respond to a situation in which $B p\sim 1-BX$ by diluting the (Einstein Frame) pressure below the threshold value or softening the spatial gradients of the scalar field. 
In this sense, the instability induced by the pressure may be analogous to the potential existence of singularities in the disformal volume element (\ref{determinant}) whenever $\bar g\propto A-2BX\to0$. This singularity in the barred metric is avoided by the field evolution, as it slows down whenever $B\dot\phi^2 \to A$. The mechanism exploited to induce a slow roll phase in cosmological applications is precisely this dynamical resistance to pathology (cf. disformal coupling to matter described in Section \ref{section:examplemodel} and disformal quintessence \cite{Koivisto:2008ak,Zumalacarregui:2010wj}).

Studying the conditions under which the pressure instability can be avoided dynamically might as well restrict the allowed functional forms of the conformal and disformal factors. In the worst case, it might spoil the disformal screening mechanism, or even completely forbid the occurrence of a disformal coupling. Determining whether or not this is the case will be the objective of future work. The Einstein Frame pressure $p$ will be neglected as a part of the approximation scheme in the following analysis, implicitly assuming that ${B p}\ll {A-2BX}$, in order to avoid potential issues.

\subsection{Coupling to Perfect Fluids} \label{perfectfluid}

Assuming a perfect fluid in the Einstein Frame $T^{\mu\nu}=(\rho+p) u^\mu u^\nu + p g^{\mu\nu}$ with $u^\al u_\al=-1$, it is instructive to project (\ref{mattereq}) along and perpendicular to the matter four velocity. This determines how the local law of energy conservation and the geodesic equation are modified by the coupling
\begin{equation}\label{perfectfl1}
u^\al\nabla_\al \rho + (\rho + p)\nabla_\al u^\al  = Q \phi_{,\al}u^\al\,. 
\end{equation}
\begin{equation}\label{perfectfl2}
(\rho + p)u^\al\nabla_\al u^\mu + \left[g^{\mu\al} + u^\mu u^\al \right] (\nabla_\al p + Q \nabla_\alpha \phi)=0\,. 
\end{equation}
In the first equation the coupling modifies the energy conservation relation, due to the energy transfered from the scalar field, which is modulated by the projection of the field gradient along the 4-velocity.
The second equation determines the departure of geodesic motion with respect to the gravitational metric. The first term describes the force arising from the pressure gradient and the second the additional force exerted by the scalar field. Both forces are projected into the direction $\perp u^\mu$ (coefficient in brackets) due to the orthogonality of the four velocity and four acceleration.

The analogue of (\ref{cov-field}) for the covariant matter conservation equation without second order field derivatives can not be obtained without choosing a time slicing due to the different high derivative structure in both equations. Nevertheless, there is no need to do so, since we already found a \emph{bona fide} field equation (\ref{cov-field}) that can be integrated consistently with the corresponding equation for matter (\ref{mattereq}), substituting the appropriate value of $Q$.
It is possible to solve for the time derivatives of all the variables after a metric ansatz has been chosen, as will be done in Section \ref{section:disfcosmo} for the study of FRW models and cosmological perturbations.

\section{The Disformal and Vainshtein Screening Mechanisms} \label{section:disfscreening}

In this Section we consider the compatibility of gravitational theories based on a disformal coupling and local gravity tests. 
Due to the stringent bounds on scalar forces and post-Newtonian effects \cite{Will:2005va,Beringer:1900zz}, some sort of screening mechanism is necessary to hide the coupling in dense environments such as the Solar System. 
The disformal contribution to the conservation equations vanishes for static, pressureless configurations \cite{Noller:2012sv}. This is obvious from (\ref{mattereq}), since only the $T^{00}$ component is nonzero for dust, and when contracting the field derivatives with the stress tensor, a non-vanishing result requires time evolution of the scalar field. 
Therefore, addressing the effects of disformal couplings requires studying the field dynamics in high density environments.%
\footnote{It is possible to obtain some insight into the dynamics of the field from the analysis of the background cosmology given in Section \ref{section:disfcosmo}, where it is argued that for the purely disformal case the coupling was proportional to the scalar field energy density, cf. Eq. (\ref{QdFRW}). This causes the existence of two regimes, a matter dominated regime in which the effects of the coupling are small, and a field dominated regime in which the coupled matter equation of state is modified, see Figure \ref{frwbackground}. When denser regions form, the scalar field energy density becomes insufficient to produce large effects on the matter distribution, \emph{unless} the field gradients follow the matter distribution and intensify the additional force.}
As it will be shown in the next subsection, the kinetic mixing induced by the disformal coupling makes the scalar field evolution insensitive to the matter distribution, as long as it is sufficiently non relativistic and its energy density is high.

For the sake of concreteness, let us restrict ourselves to a canonical scalar field coupled to a perfect fluid. 
The general equation (\ref{cov-field}) then reads
\begin{eqnarray}\label{covfield-canonical}
\lp \mathcal X g^{\mu\nu} - B\, T^{\mu\nu}_{m} \rp \nabla_\mu\nabla_\nu\phi - \mathcal X V' + \mc Q_{\mu\nu} T^{\mu\nu}_{m}  
=0 \,,
\end{eqnarray}
where $\mathcal X \equiv  A-2BX < A$ is bounded in order to avoid a singularity in the volume element of the barred metric (\ref{determinant}) and $\mc Q_{\mu\nu}$ is given by Eq. (\ref{Qpot}). 
The form of the field equation strongly suggests that the dynamics of the coupled system will be different in high than in low density environments. Since the energy momentum-tensor appears as a coefficient of the higher derivatives as well as in the effective potential, there is a well defined limit $T^{00}=\rho\to\infty$, in which the field equation simplifies considerably. This property will be crucial for the \emph{disformal screening mechanism} \cite{Koivisto:2012za}.
Additionally, the study of static vacuum configuration around point sources allows one to derive the existence of a \emph{Vainshtein radius}, at which the asymptotic solution $\propto r^{-1}$ breaks down. This property arises from a singularity in $\mc X \equiv  A-2BX $ for sufficiently high field gradients, whenever $B<0$. The disformal screening mechanism, the existence of the Vainshtein radius and potential signatures of disformally coupled theories will be analyzed in below.

\subsection{Dense, Non-relativistic and Static Matter}\label{fielddecoupl} 

The study of Solar System and laboratory tests of gravity requires considering energy densities that are much higher than the cosmological average and pressure is completely sub-dominant.
As a first approximation, this regime can be explored using the general scalar field equation (\ref{covfield-canonical})
for a static, non-relativistic matter distribution $\rho(\vec x)$ in the limit $\rho\to\infty$. More precisely, the following dimensionless ratios will be assumed to be negligible
\begin{equation} \label{approximations-screening}
\frac{p}{\rho}\,,\; 
\frac{ p}{\rho} \lp \frac{\vec \partial\phi}{\partial_t\phi}\rp^2  \,,\; 
\frac{\mathcal X}{B\rho} \,,\;
\frac{\mathcal X}{B\rho} V'/\ddot\phi \,,\;
\Gamma^\mu_{00}\phi_{,\mu} /\ddot\phi
\; \sim 0\,,
\end{equation}
and $Bp<A-2BX$, as argued in Section \ref{Bp_instab}.
These approximations quantify to what extent the effects of gravity and pressure are disregarded, the requirement of having ``soft'' spatial gradients relative to the time evolution and the fact that $B\rho$ is large enough. These conditions will be briefly discussed at the end of the Section, focusing on systems in which their lack of fulfillment might lead to observable signatures.

The set of assumptions (\ref{approximations-screening}) simplifies the field equation (\ref{covfield-canonical}) considerably
\begin{equation} \label{time}
\ddot{\phi} \approx -\frac{B'}{2B}\dot{\phi}^2+A'\lp\frac{\dot{\phi}^2}{A}-\frac{1}{2B}\rp 
= -\frac{\bt}{2M_p}\dot{\phi^2}\,,
\end{equation}
where the first equality is general and the second applies to a purely disformal coupling with exponential forms, such as the example cosmological model presented in Section \ref{section:examplemodel}.
The above expression departs substantially from the simple conformal coupling, for which the $\rho\to\infty$ limit is ill-defined. Two important features of the above equation endow the theory with the \emph{disformal screening mechanism}:
\begin{itemize}
 \item The spatial derivatives become irrelevant, as they are suppressed by a $p/\rho$ factor with respect to the time derivatives.%
\footnote{Equation (\ref{time}) also follows from taking the limit $\rho\gg A/B,\dot\phi^2$ in the FRW coupling density $Q_0$ (\ref{kg}), precisely due to the absence of spatial derivatives.}
\item The equation becomes {\it independent of the local energy density}, making the field evolution insensitive to the presence and distribution of massive bodies. 
\end{itemize}
These features ensure that \emph{the field rolls homogeneously} and avoids the formation of spatial gradients between separate objects, which would give rise to the scalar force (cf. Section \ref{section_pointparticle} ). The above properties are caused by the kinetic mixing between the field and matter degrees of freedom, and lay at the core of the decoupling between both components.

Let us analyze the simpler, purely disformal exponential case. The second equality of equation (\ref{time}) can be easily integrated
\begin{equation}\label{time2}
 \dot\phi(t)=\frac{M_p}{\beta}  \left(t + \frac{M_p}{\beta\dot\phi(0)} \right)^{-1}  \,.
\end{equation}
In this solution the field time variation is approximately constant while $t \ll \frac{M_p}{\beta\dot\phi(0)}$ and slows down afterwards as $\propto 1/t$. Since the coupling to non-relativistic matter is proportional to $\dot\phi$, stronger couplings decay earlier. 
It is possible to obtain a solution for $A=1,\, \dot\rho =0$ keeping the potential $V$, but otherwise assuming the simplifications (\ref{approximations-screening}).
It is given as an implicit function
\begin{equation} \label{timeV}
 t - t_0 = \int_{\phi_0}^{\phi}\sqrt{\frac{B(\phi')}{C_0 -2V(\phi')/\rho}}d\phi'\,,
\end{equation}
where $\dot\phi^2 = \frac{C_0}{B} - \frac{2V(\phi)}{B\rho}$.%
\footnote{Under these assumptions, Eq. (\ref{covfield-canonical}) can be written as
$\ddot\phi + \frac{B'}{2B}\dot\phi^2 + \frac{V'}{B\rho} 
= \frac{1}{B\dot\phi}\frac{d}{dt}\lp B\dot\phi^2/2 + \frac{V}{\rho}\rp =0$.
The second equality can be directly integrated, giving the constraint
$\dot\phi^2 = \frac{C_0}{B} - \frac{2V(\phi)}{B\rho}$, which can be integrated again to obtain (\ref{timeV}).}
The potential appears suppressed with respect to the energy density. In tracking Dark Energy models, such as the one explored in Section \ref{section:examplemodel}, $V$ is a decreasing function of the field and $\dot\phi>0$. Therefore, $V$ is of the order of magnitude of the average cosmic density $\rho_0$ and can be safely neglected if $\rho$ is much higher, recovering the simpler solution (\ref{time2}). As the field slows down with time if $B'/B>0$, the order of magnitude of the field time derivative is also cosmological, $\dot\phi\sim H_0$.

One of the effects of the coupling is to modify the energy conservation equation for matter in the Einstein Frame, cf. Eq. (\ref{perfectfl1}), inducing a variation of the gravitational mass. In a gravitationally bound two body system, this effect is degenerate with a possible time evolution of Newton's constant $\frac{\dot G}{G} \leftrightarrow \frac{\dot M}{M} + \frac{\dot m}{m}$ to a first approximation, as can be argued by deriving the expression for the Newtoninan force with respect to time. 
Lunar laser ranging measurements place precise bounds on this effect to the level of $\dot G/G < 10^{-3}/{\rm Gy}$ \cite{Williams:2004qba}.  
The magnitude of energy density variation induced by a disformal coupling can be estimated as $\dot\rho\approx-(\ddot \phi + V')\dot\phi$.
Assuming $\dot\phi^2 \sim V \sim \rho_0$ as discussed above, $B\rho_0 \gtrsim 1$%
\footnote{This assumptions are made for order of magnitude estimates. For cosmological applications in which the field's potential accelerates the Universe, $\rho,V$ are significantly larger than $\dot\phi^2$ and $B\rho> 1$ occurs while $B\dot\phi^2<1$, cf. Figure \ref{disffactors}.}
and the solution (\ref{time2}), typical mass variation rates $\dot M/M$ 
are as small as $\sim 10^{-6}/{\rm Gy}$ for the interstellar medium and $\sim 10^{-29}/{\rm Gy}$ for the average Earth density, well beyond the sensitivity of Lunar laser ranging measurements.

\subsection{Static Field in Vacuum}
\label{section:vainshtein}

The equivalence between disformally coupled theories and quartic DBI Galileons presented in Section \ref{section:confframe} suggests a relation between the disformal and the  Vainshtein screening mechanisms. The Vainshtein mechanism, which occurs in Galileon theories and their generalizations, is due to the non-linear, derivative self-interactions of the scalar field, which suppress the field gradients within a certain distance from point sources in static configurations. Such a distance is known as the Vainshtein radius $r_V$, and it signals the breakdown of the asymptotic vacuum solution as the non-linear terms become dominant.

Let us consider static and spherically symmetric solutions of the scalar field equation (\ref{covfield-canonical}). For perfect vacuum $T^{\mu\nu}_m=0$, it reduces to the Klein-Gordon equation in the vacuum: $\Box \phi -V' = 0$. For the sake of concreteness, we will assume that $V=0$ throughout this Section. Then the vacuum solution is
\begin{equation}\label{vacuumphi}
\phi(r)= -\frac{S}{r} + \phi_0\,.
\end{equation}
If the field is conformally coupled to matter, such a static field configuration would produce an additional force, which is unscreened (screened) if  $A'|S|\gtrsim GM$ ($A'|S|\ll GM$) in the regime of validity of the above solution. 

The Vainshtein radius can be found by determining the breakdown of the conditions that lead to solution (\ref{vacuumphi}). In order to do so, let us consider a very small, but non zero ambient energy density such that $B\delta T^{\mu\nu}_m \phi_{;\mu\nu}\ll \Box \phi$.
Then the field equation (\ref{cov-field}) reduces to
\begin{equation} \label{vainshtein}
 \Box \phi + \lp A+2B(\phi_{,r})^2 \rp^{-1} \mc  Q_{\mu\nu} \delta T^{\mu\nu}_m  \approx 0 \,.
\end{equation}
The vacuum solution satisfies $\Box\phi=0$, ad hence the above equation is only satisfied as long as the second term remains small. But if solution (\ref{vacuumphi}) is assumed and $B<0$, the coefficient in parenthesis has a singularity at
\begin{equation}
r_V = \left( \frac{2 B S^2}{A}\right)^{1/4}\,.
\end{equation}
We can identify $r_V$ as the Vainshtein radius, as it determines the breakdown of the na\"ive vacuum solution, which is independent of $\delta T^{\mu\nu}_m$ (note that an experimental set-up able to measure the scalar force would generically introduce such a non-vanishing energy content).%
\footnote{It is natural to include a non-vanishing energy component to regularize the equations: in the absence of a matter Lagrangian, the results of Section \ref{section:confframe} imply that a theory described by a quartic DBI Galileon (\ref{LagGF2}) is equivalent to General Relativity, as the scalar field can be eliminated by a redefinition of the metric (the canonical kinetic term for the scalar field would transform into a disformal quintessence term (\ref{disfquint})).}
This is analogous to the Vainshtein radius in the Jordan or Galileon frames, at which the non-linear derivative self-interactions of the scalar field become important.

Disformally coupled theories do therefore display the Vainshtein effect in static, vacuum configuration for negative $B$, while they are endowed with the disformal screening mechanism for positive $B$. It is worth noticing that both effects are essentially different: the disformal screening relies on the kinetic mixing of the matter and the field, which can render the field evolution independent of the energy density. When regarded in the Einstein frame, the Vainshtein mechanism is related to a singularity in the barred metric, as $\bar g\propto A-2BX \approx A - B\dot\phi^2 + B(\vec\nabla\phi)^2$ approaches zero.%
\footnote{This is similar to the mechanism to induce slow roll through a disformal coupling, cf. Section \ref{section:examplemodel} or self-coupling \cite{Koivisto:2008ak,Zumalacarregui:2010wj}.}
Moreover, the occurrence of each mechanism required a different sign of the disformal coupling function $B$: $B>0$ gives rise to the disformal screening mechanism while $B<0$ is related to the Vainshtein effect. In both cases the relevant energy scale for the screening is given by $B\sim M^{-4}$, and in particular by the dimensionless quantities $B\rho$ (Disformal) and $B(\vec\nabla\phi)^2/A$ (Vainshtein).
The relationship between the two screening mechanisms, as well as a more detailed comparison with other theories that feature the Vainshtein screening will be investigated in future work.

\subsection{Potential Signatures}\label{s:signatures}

It has been shown that modifications of gravity might be rendered small in the Solar System by the action of the disformal and Vainshtein screening mechanism. New local, astrophysical and cosmological signatures may be found by relaxing the approximations assumed in the previous sections. Some situations where the coupling might become observable include:

\setcounter{paragraph}{0}

\paragraph{Matter velocity flows:} The spatial component of the matter four-velocity $T^{0i}$ mixes the time and space derivatives of the field, which may source the field evolution (as non-zero velocities introduce terms proporional to $\phi_{;0i}$ and $\dot\phi\,\phi_{,i}$). These effects are suppressed by a relativistic $v/c$ factor, but they may be important in certain systems such as binary pulsars.

\paragraph{Pressure:}  Applications of the disformal coupling in the context of Dark Energy arguably require a value $B\rho_0\gg 1$, where $\rho_0$ is the average cosmic density. Then, even though the pressure is usually negligible with respect to the energy density, it should be easy to find systems for which $Bp$ is also much larger than one. This might have important consequences for the stability of the theory, as was briefly discussed in Section \ref{Bp_instab}.
                                                                                                                                                                                                                                                                                                                                                                                                                                                                                                                                                                                                                                             
\paragraph{Radiation:} Unlike in the conformal case, the disformal coupling has non-trivial effects on ultra-relativistic fields for which $T\approx 0$, cf. (\ref{p_div}). Some authors have initiated the study of the disformal coupling in scenarios featuring radiation. Brax \emph{et al.} \cite{Brax:2012ie} considered high-precision, low-energy photon experiments, which might be able to detect the influence of a disformal coupling on top of a conformal one. The distortions in the baryon-photon chemical potential induced by a disformal coupling and their signatures on the CMB small scale spectrum have been studied by van de Bruck and Sculthorpe \cite{vandeBruck:2012vq}.
Other effects may follow if Electromagnetism is formulated in terms of the barred metric, such as varying speed of light or modified gravitational light deflection \cite{Wyman:2011mp}. A radiation-exclusive disformal coupling has also been shown to affect the evolution of the CMB temperature \cite{vandeBruck:2013yxa}.

\paragraph{Strong gravitational fields:} The connection coefficient $\Gamma^{\mu}_{00}\phi_{,\mu}$ in the field derivative term is not suppressed by $B\rho$. It  represents the effects of gravity, and was neglected because it is small in most Solar System applications, since $\Gamma^{r}_{00}=\frac{GM}{r^3}(r-2GM)$  in the Schwarzschild metric. However, this term might become relevant in strong gravitational fields, such as the vicinity of black holes or compact objects.

\paragraph{Spatial Field Gradients:} 
In the $B\rho\gg1$, $\rho\gg p$ limit, the equation for the scalar field (\ref{time}) becomes independent of the matter content and the field derivatives. Therefore, if the field acquires a spatial modulation before reaching this limit, it will be preserved by the subsequent evolution. 
Spatial gradients of the field formed when the linear perturbation theory is valid would then be present today, with their actual value depending on the details of the transition between the perturbative, e.g. the small scale limit of cosmological perturbations (\ref{disfcoup:smallscale}), and the screened regimes. Gradients of cosmological origin might be seen as preferred direction effects pointing towards cosmic structures when analyzed in the Solar System.
Spatial derivatives of the field may also be important if the field is rolling sufficiently slow as to overcome the $p/\rho$ factor in (\ref{approximations-screening}).

These and other settings might lead to characteristic signatures and new bounds for disformally coupled theories, which will be investigated in the future. It should be also possible to obtain the coefficients of the Parameterized Post Newtonian approximation, which would allow a more systematic comparison to local gravity tests.


\section{Cosmology} \label{section:disfcosmo}

Having addressed the viability of the theory in the Solar System, let us consider its cosmological implications.
Using the Einstein Frame description, the Friedmann equations have the usual form
\begin{eqnarray}
 H^2 + \frac{k}{a^2}  &=&  \frac{8\pi G}{3}( \rho + \frac{\dot{\phi}^2}{2}+V)\,,   \\
\dot{H}+H^2 &=&  -\frac{4\pi G }{3} ( \rho + 2\dot{\phi}^2-2V)\,, 
\end{eqnarray}
but the conservation equations for matter and the scalar field have to be computed from (\ref{mattereq}, \ref{cov-field}):
\begin{eqnarray} \label{background-dm}
&& \dot{\rho}+3H\rho =  Q_0\dot{\phi} \,, \\
&& \ddot{\phi}+3H\dot{\phi}+V'  =  -Q_0\,, \label{kg}
\end{eqnarray} 
were  $\rho$ is the energy density of the coupled matter component and the background coupling factor reads
\begin{equation}\label{background-coupling}
Q_0 = \frac{A'-2B(3H\dot{\phi}+V'+\frac{A'}{A}\dot{\phi}^2)+B'\dot{\phi}^2}{2\lp A+B(\rho-\dot{\phi}^2)\rp}\rho\,,
\end{equation}
after solving away the higher derivatives. In the following we restrict to flat space, $K=0$.

At this stage it is possible to understand the difference between the pure conformal ($B=0$) and disformal ($A=1$) cases by writing (\ref{background-coupling}) in terms of the equation of state and the scalar field energy density:
\begin{equation}
 Q_0^{\rm (c)} = \frac{A'}{2A} \rho\,, \hspace{.6cm} 
\end{equation}
\begin{eqnarray}
Q_0^{\rm (d)} &\approx & \Big( \frac{B'}{2B}(1+w_\phi) - \frac{V'}{2V}(1-w_\phi) \Big) \rho_\phi \nonumber  \\
&& + \frac{\sqrt{3}}{M_p}\lp (1+w_\phi)\rho_{\rm tot}\rho_\phi\rp^{1/2}\,,\label{QdFRW}
\end{eqnarray}
where in the pure disformal case it has been assumed that $B\rho\gg 1 \gtrsim B\dot\phi^2$. This approximation is satisfied by the model presented in the next subsection when the coupling is active, see Figure \ref{disffactors}. The last term in (\ref{QdFRW}) represents the contribution from the Hubble term, which is subdominant when the slopes of $B,V$ are large. The above expressions imply that the conformal and disformal coupling between Dark Energy and Dark Matter are related to essentially different phenomenological parameterizations, where the interaction is either proportional to $\rho=\rho_{\rm dm}$ \cite{Casas:1991ky,Amendola:1999er,Koivisto:2005nr} or $\rho_{\phi}$ \cite{Clemson:2011an,Chen:2011cy}. 

The equations governing cosmological perturbations can be obtained from Eq. (\ref{divergence}), which can be used to read both the disformal matter non-conservation and the field dynamical equation. Working in the Newtonian gauge%
\begin{equation}
 ds^2=-(1+2\Phi)dt^2 + a^2(1-2\Psi)d\vec x^2 \,,
\end{equation}
avoids potential misinterpretations when swapping between different frames, at least when these are related by conformal transformations \cite{Brown:2011eh}. 
Solving for the higher order derivatives, the perturbed continuity and Euler equations for the disformally coupled matter contrast $\delta_{\rm dc}=\delta\rho/\rho$ and the divergence of its velocity $\theta=ik_jT^{0i}_{(m)}=ik_j v^j \rho a^{-1}$ read
\begin{eqnarray}\label{cp}
\dot \delta_{\rm dc} + \frac{\theta}{a} +\frac{Q_0}{\rho}\dot\phi \delta_{\rm dc}
&=& 3 \dot \Psi + \frac{Q_0}{\rho} \delta\dot \phi + \frac{\delta Q}{\rho} \dot \phi \,, \\
\dot \theta + \theta\left(H + \frac{Q_0}{\rho} \dot\phi\right)  &=&  k^2 \lp\Phi + \frac{Q_0}{\rho} \delta\phi\rp \,, \label{euler}
\end{eqnarray}
while the scalar field evolution is determined by
\begin{widetext}
\begin{eqnarray}
\delta \ddot \phi + 3  H \delta \dot \phi + \left( \frac{k^2}{a^2} + V'' \right) \delta \phi 
  =   -\delta Q - 2\Phi \left(Q_0 + V'\right)
+  \dot \phi( \dot \Phi +3 \dot \Psi ) \label{kgp} \,,
\end{eqnarray}
with
\begin{eqnarray}
\delta Q^{\rm (d)} & = & -\left(\frac{k^2}{a^2}\frac{B\rho}{M}
+(2 B V'' -B''\dot\phi^2 )\frac{\rho}{2M}
+  \left(2 B'(V'+3 H \dot \phi) 
+ B'^2 \dot\phi^2 (\rho-\dot \phi^2)\right)\frac{\rho}{2M^2}
 \right)\delta\phi \nonumber
\\ 
&&  
+ (1-B\dot\phi^2) \frac{Q_0}{M}\delta_{\rm dc} 
+ \left(B' \dot\phi - B (3 H-\rho B' \dot\phi)
-B^2 (2 V'\dot \phi+3 H(\rho+\dot\phi^2))\right)\frac{\rho}{M^2}\delta\dot\phi 
 \nonumber
\\
&&
+ \left(-B' \dot \phi + B (6 H - \rho B' \dot \phi )
+ 2 B^2 (3 H \rho+V' \dot \phi )\right) \frac{\rho \dot\phi}{M^2} \Phi
+  \frac{3 B \rho \dot\phi }{M}\dot\Psi \,,
\label{perturb-disformal}
\end{eqnarray}
\end{widetext}
for a purely disformal coupling $A=1$, where $M = 1+B \rho -B \dot\phi^2$. The coupling perturbation $\delta Q$ is given in Appendix \ref{generalDelQ} for the general case. This expression is a much more cumbersome combination of the fluid and field perturbations than for the purely conformally coupled case
\begin{equation}
\delta Q^{\rm (c)} = \frac{1}{2} \log(A)'\rho\delta_{\rm dc} + \frac{1}{2}\log(A)''\rho\delta \phi\,. 
\end{equation}
Note that, unlike in the conformal case, the first term in (\ref{perturb-disformal}) is proportional to $k^2$ and hence the coupling introduces explicit scale dependent terms at the level of the equations. This feature will be reflected in the growth of perturbations and the power spectrum, studied below for an example model.

To extract the most relevant new features by analytic means, we shall consider the subhorizon approximation.
In the small scale limit, taking into account only the matter perturbations and the gradients of the field, there is a simple expression for the perturbed interaction $\delta Q$. In this Newtonian limit, we further relate the field gradient to the matter perturbation through the field equation  (\ref{kgp}), which yields the simple expression 
\begin{equation}
 \delta  Q^{\rm (N)} = Q_0\delta_{\rm dc}\,.
\end{equation}
Combining equations (\ref{cp}) and (\ref{euler}) together with the usual Poisson equation, we obtain the evolution of the coupled Dark Matter overdensity   
\begin{equation}
\ddot{\delta}_{\rm dc}+\lb 2H+\frac{Q_0}{\rho}\dot{\phi}\rb \dot{\delta}_{\rm dc} = 4\pi G_{\rm eff}\rho\delta_{\rm dc}\,.
\end{equation} 
In addition to an extra friction term, the source term is modulated. The last effect is captured by defining an effective gravitational constant $G_{\rm eff}$ that determines the clustering of Dark Matter particles on subhorizon scales
\begin{equation} \label{disfcoup:smallscale}
\frac{G_{\rm eff}}{G}-1=\frac{Q_0^2}{4\pi G\rho^2}\,. 
\end{equation} 
This approximation has the same expression as the simple conformal case, although with a significantly different functional form of the coupling $Q_0$, which is now given by Eq. (\ref{background-coupling}).

\subsection{Disformally Coupled Dark Matter}\label{section:examplemodel}

In what follows it will be assumed that the field is only coupled to Dark Matter, while radiation and baryons follow geodesics of the gravitational metric and do not feel the scalar interaction directly - If baryons are also coupled, then the ratio $\rho_{dm}/\rho_b$ remains fixed, because both species feel the same effective metric.%
\footnote{This can also be seen directly from (\ref{background-coupling}): in the denominator of $Q_0$, the energy density has to be substituted by the total one $\rho\to \rho_{dm}+\rho_b$, while the multiplicative coefficient $\rho$  would refer to each individual species.}
Postulating that the baryonic and electromagnetic sectors are constructed out of the gravitational metric also avoids problems with precision gravity tests and the subtleties related to the existence of different frames, hence simplifying the analysis of cosmological observations.

To study the dynamics within a particular example, we focus on a simple \emph{Disformally Coupled Dark Matter} (DCDM) model, constructed with the following prescriptions:
\begin{itemize}
 \item Dark Matter disformally coupled to a canonical scalar field, following Eq. (\ref{background-dm}-\ref{background-coupling}).

\item An exponential parametrization for the disformal relation and the scalar field potential:
\begin{eqnarray} \label{examplemodel}
B &=& B_0 e^{\beta(\phi-\phi_0)/M_p}\,, \label{examplemodel2} \\
V &=& V_0 e^{-\gamma\phi/M_p}  \,,  \\
A &=& 1 \label{examplemodel3} \,. 
\end{eqnarray}
with $M_p=(8\pi G)^{-1/2}$. The conformal factor $A$ has been set to the trivial value in order to focus on the novel features. Furthermore, the coupling is chosen to be negligible in the early universe, and hence initial conditions and early evolution are not affected.

\item Uncoupled baryons, photons and neutrinos, which follow the usual barotropic scaling relations $\rho=a^{3(1+w)}$. Zero cosmological constant.
\end{itemize}

This model can be motivated in DBI scenarios in Type IIB string theory, in which Dark Matter is given by the fields residing on the brane, which is allowed to move in the compact dimensions, being then automatically dark and disformally coupled \cite{DBIpaper}.
Besides being motivated from some high energy scenarios, the exponential forms (\ref{examplemodel}-\ref{examplemodel3}) facilitate the choice of natural scales for the constant prefactors by shifting the zero point of the field (e.g. $B_0\sim M_p^{-4}$, $V_0\sim M_p^{4}$, $A_0$ dimensionless). Furthermore, these forms allow a convenient exploration of the phase space of the system. In addition to the previously studied fixed points \cite{Amendola:1999er,Copeland:2006wr}, we find only one new, a disformal scaling solution that is not an attractor.  The details of this analysis can be found in Appendix \ref{dynamicalsystem}.

\subsubsection{Background Evolution}

\begin{figure}
\includegraphics[width=\columnwidth]{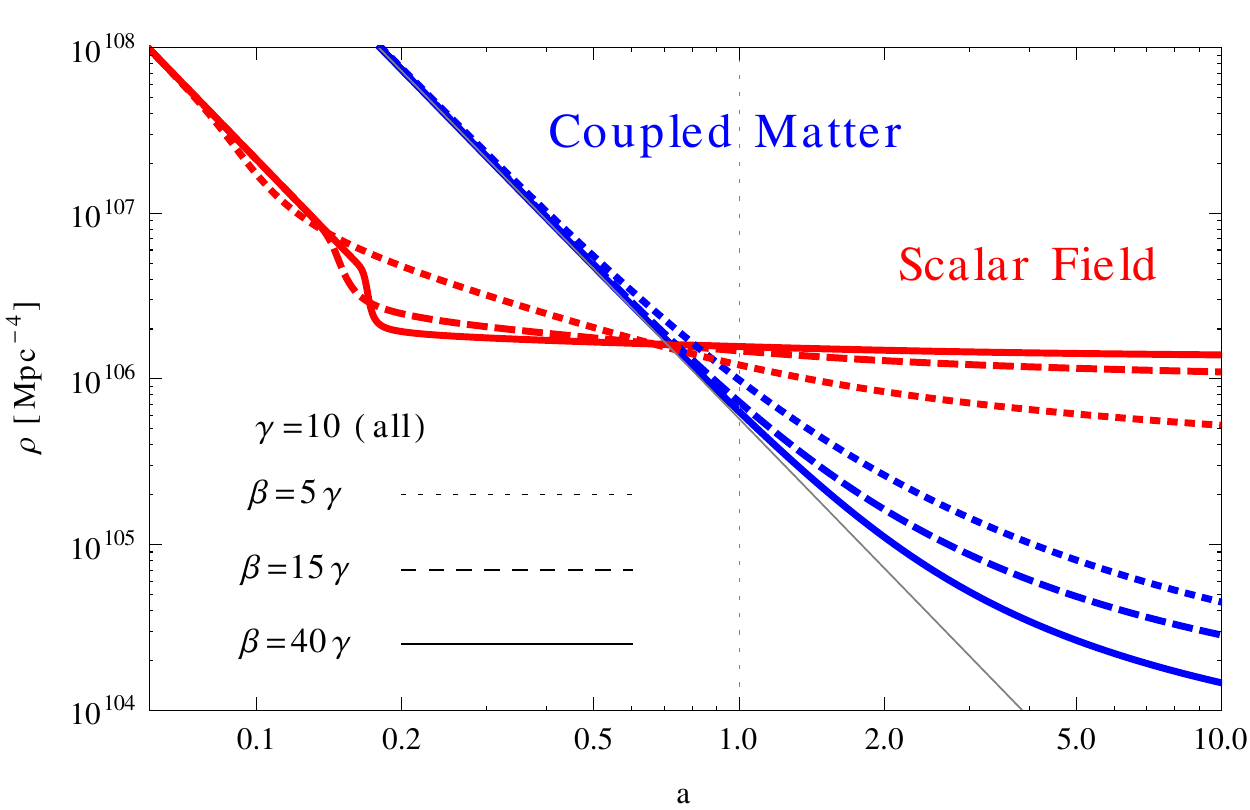}
\includegraphics[width=\columnwidth]{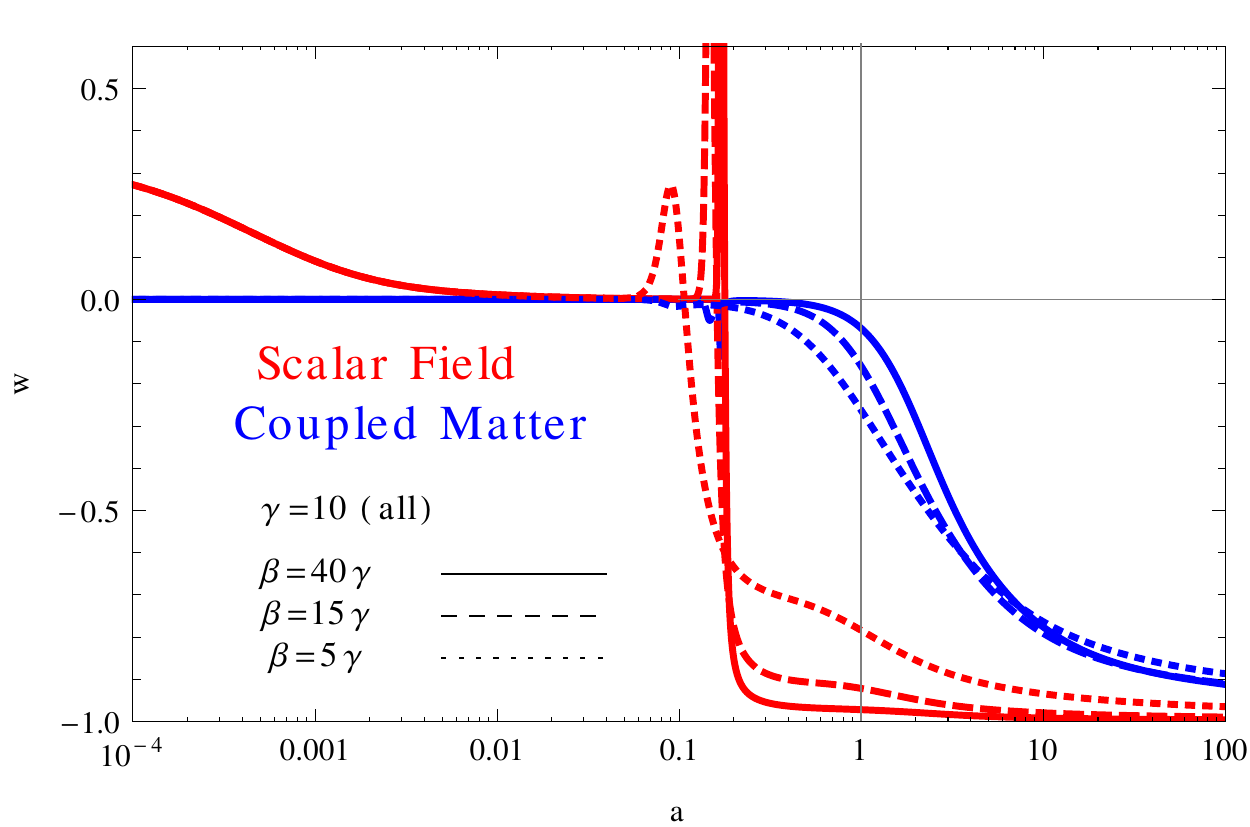} 
\caption[Background evolution of disformally coupled matter]{ \justifying
Background evolution of disformally coupled matter. {\bf Left:} evolution of the energy density for the field (red, light) and coupled matter (blue, dark) for different choices of the coupling slope $\beta$.
{\bf Right:} equation of state for the field (red, light) and coupled matter (blue, dark). High values of $\beta/\gamma$ (solid, dashed) give a good fit to observations, while low values (dotted) do not produce enough acceleration.
\label{frwbackground}}
 \end{figure}

\begin{figure}
\includegraphics[width=\columnwidth]{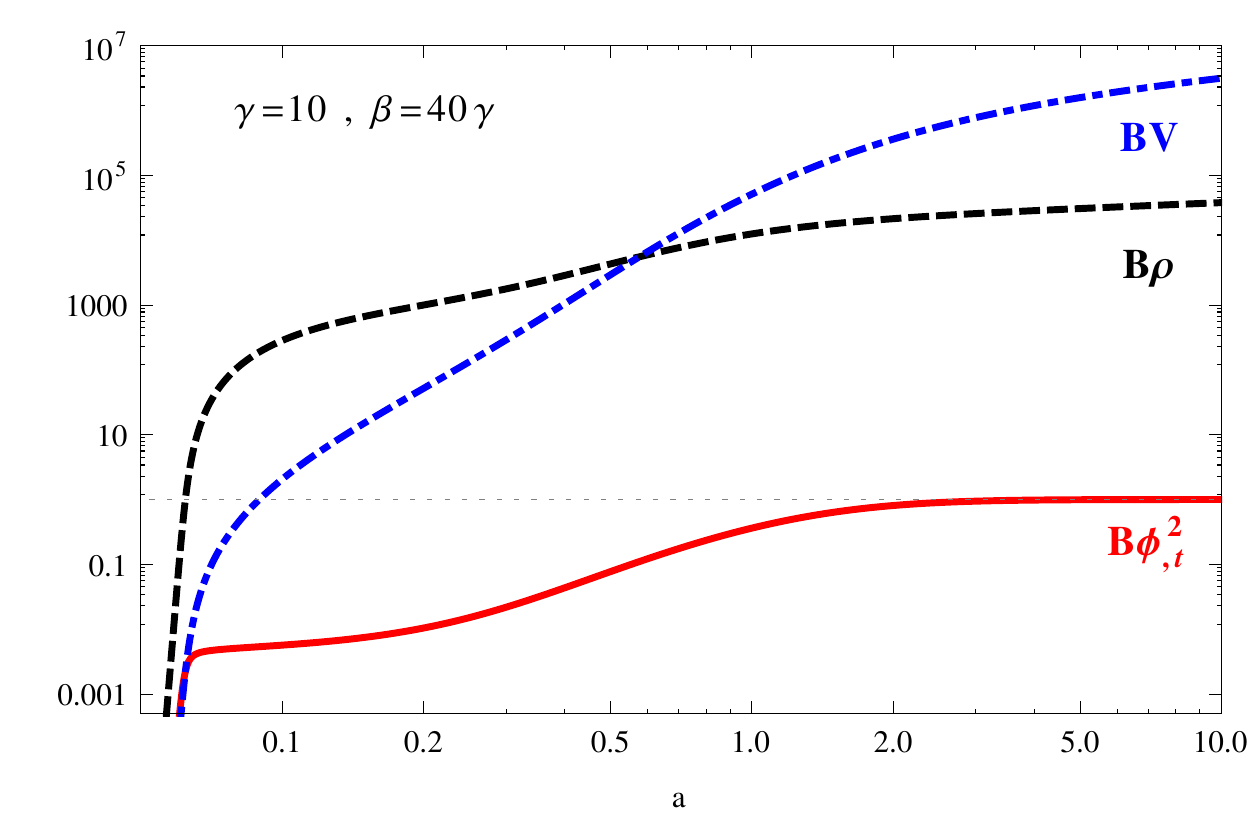}
\includegraphics[width=\columnwidth]{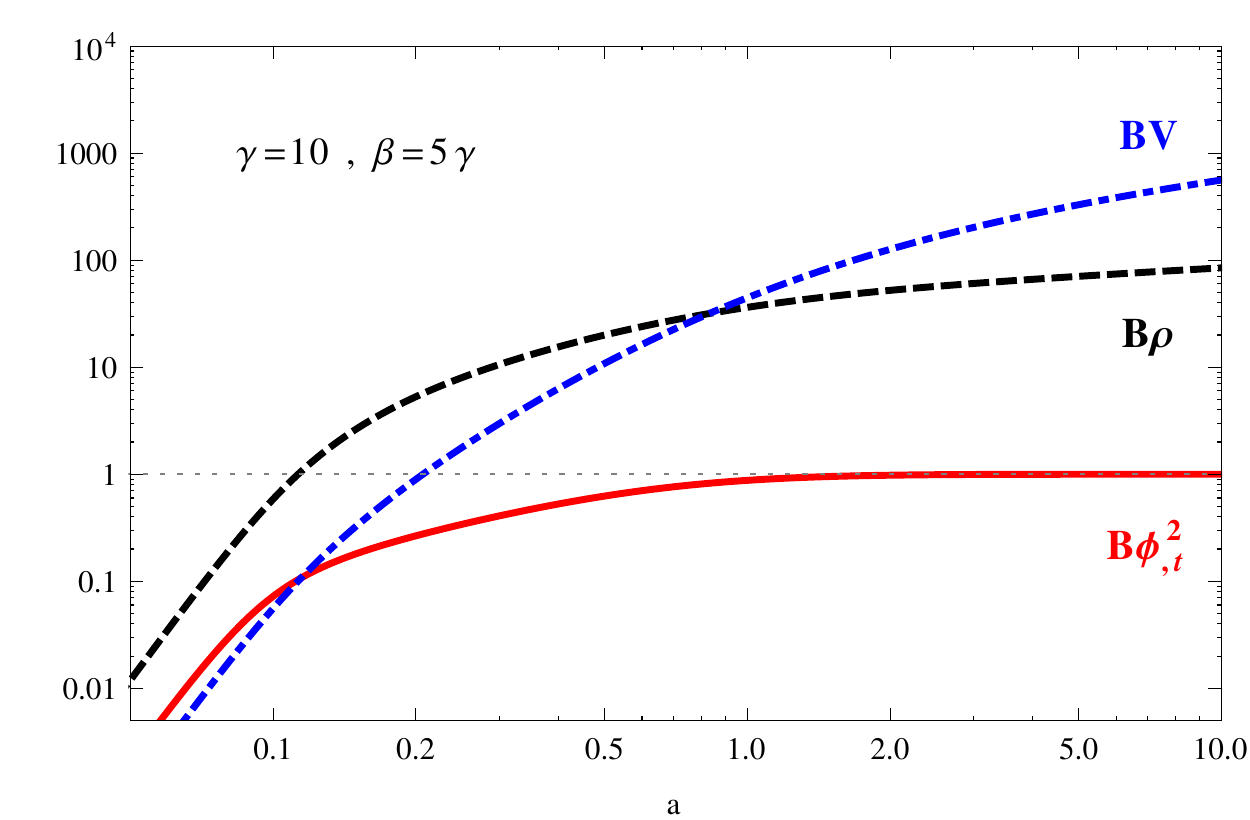}
\caption[Evolution of the dimensionless disformal factors]{ \justifying
Evolution of the dimensionless disformal factors $B\dot\phi^2$ (solid), $B\rho$ (dashed) and $BV$ (dot-dashed). Higher values of $\beta/\alpha$ produce a sharper transition and lead to higher $B$ at late times.
\label{disffactors}}
\end{figure}

The model set up is similar to the uncoupled self-interacting field case described in Ref. \cite{Zumalacarregui:2010wj}. In particular, the potential ensures a tracking stage for the field and the value of $\phi_0$ is chosen to tune the transition time when the disformal coupling becomes relevant.
Although only Dark Matter is affected by the coupling, radiation and baryons are included in order to provide a more realistic description.
The evolution at early times is then as in the usual exponential quintessence model, where the field tracks the dominant fluid component and the slope of the potential $\gamma$ determines the amount of Early Dark Energy (EDE) \cite{Copeland:2006wr}
\begin{equation}\label{Om_ede}
 \Omega_{\rm ede} = \frac{3}{\gamma^2}(1+w_m)\,,
\end{equation}
which depends on the dominant matter component equation of state parameter $w_m$.
The new features appear when the disformal factor $B\dot\phi^2$ becomes of order one. Then the clocks that tick for Dark Matter, $\bar{g}_{00} = -1+B\dot{\phi}^2$, slow down and make the effective equation of state approach minus unity asymptotically. The field also slows down to avoid a singularity in the effective metric $\bar{g}_{\mu\nu}$, and the universe enters into a de Sitter stage. This natural resistance to pathology was also observed in the disformal self-coupling scenario described in Refs. \cite{Koivisto:2008ak,Zumalacarregui:2010wj}. 
The disformal coupling provides then a mechanism that triggers the transition to an accelerated expansion. The relatively steeper the slope of the disformal function is, i.e. the higher the ratio $\beta/\gamma$, the faster the transition happens, as seen in Figures \ref{frwbackground}, \ref{disffactors}. This transition also produces a short ``bump'' in the equation of state, which affects the growth of structure.

\begin{figure}\vspace{15pt}
 \begin{center}
 \includegraphics[width=\columnwidth]{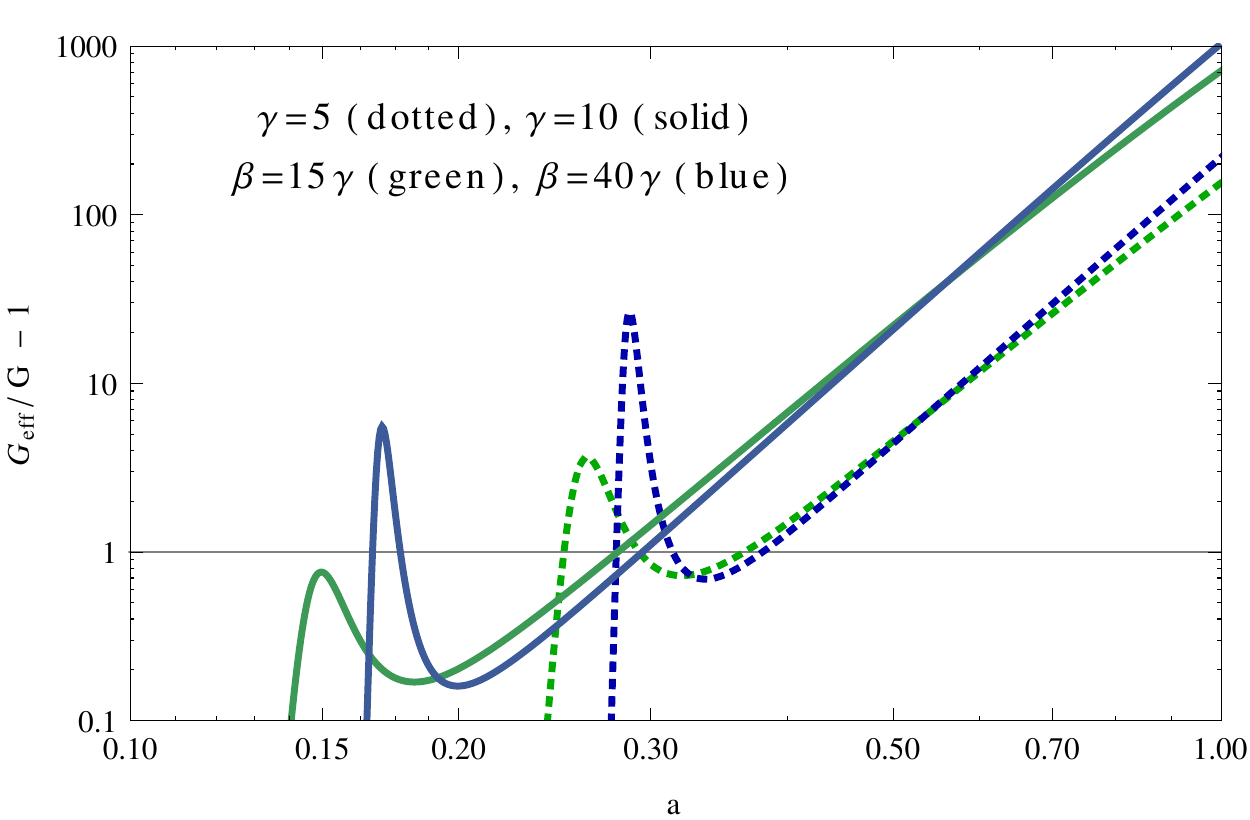}
\end{center}
\caption[Effective gravitational constant for disformally coupled matter]{\justifying
Effective gravitational constant on small scales (\ref{disfcoup:smallscale}) for different values of $\beta,\gamma$. The value is large at the transition due to the disformal friction term $B'\dot\phi^2$, and latter due to the contribution of the potential term $BV'$ (see text and compare to Figure \ref{frwbackground}).
\label{Geff}}
\end{figure}

%
The evolution of $G_ {\rm eff}$ for the disformally coupled Dark Matter example model (\ref{examplemodel}-\ref{examplemodel3}) is shown in Figure \ref{Geff}. It is characterized by a bump at the transition, whose height increases with $\beta$, and a further increase when the potential becomes dominant. At the later stage, the dependence is approximately $G_{\rm eff}/G -1 \sim \lp \gamma V/\rho\rp^2$ and yields a large value since Dark Energy domination requires $V\gtrsim\rho$ and  $\gamma\gtrsim 15$ is necessary to avoid the effects of early Dark Energy (\ref{Om_ede}) \cite{Reichardt:2011fv}.
This enhancement occurs on observable scales and spoils the formation of large scale structure in this particular case. Problematic growth enhancement also occurs in conformally coupled models that attempt to address the coincidence problem \cite{Koivisto:2005nr}. The observable effects will be analyzed using the full perturbation $\delta Q$ within the disformally coupled Dark Matter model. Several alternatives to render the model viable will be described in Section \ref{section:disfviable}.

\subsubsection{Perturbations}

\begin{figure*}
 \begin{center}
 \includegraphics[width=\columnwidth]{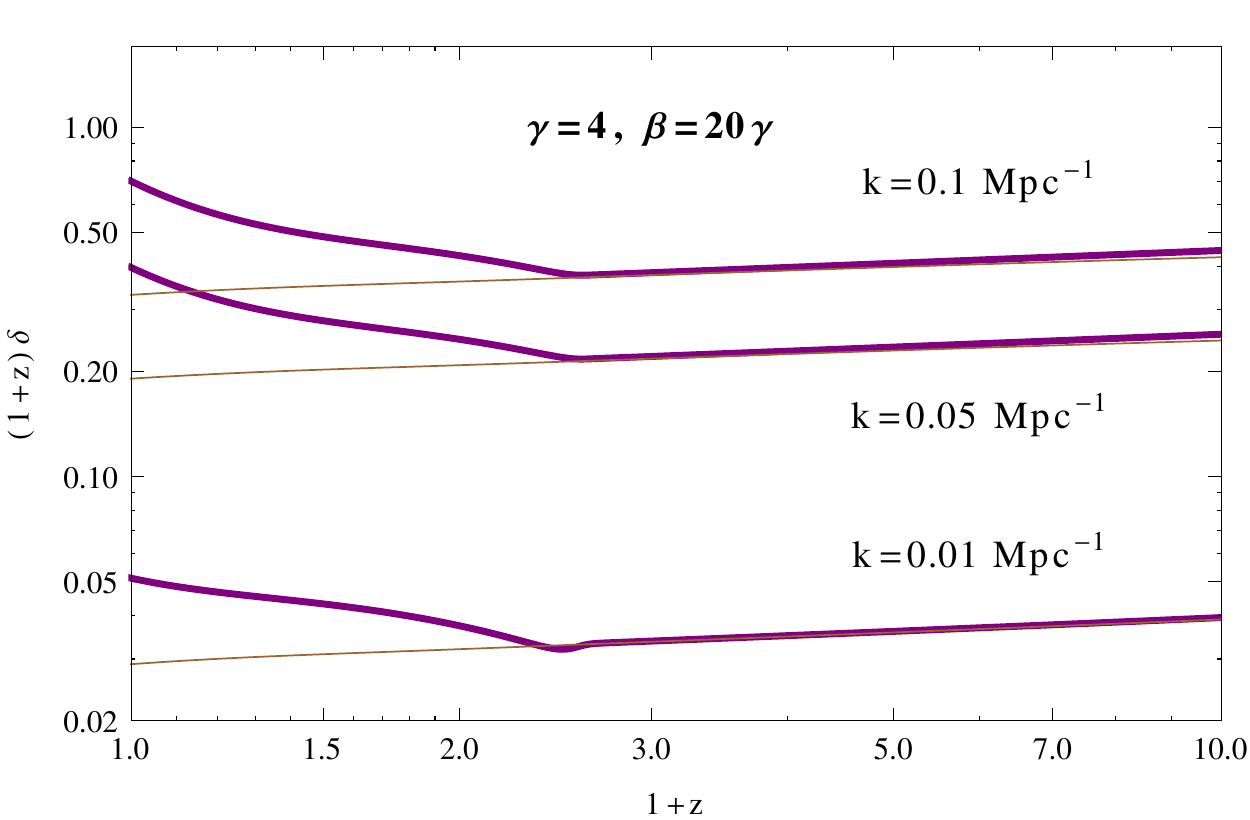}
 \includegraphics[width=\columnwidth]{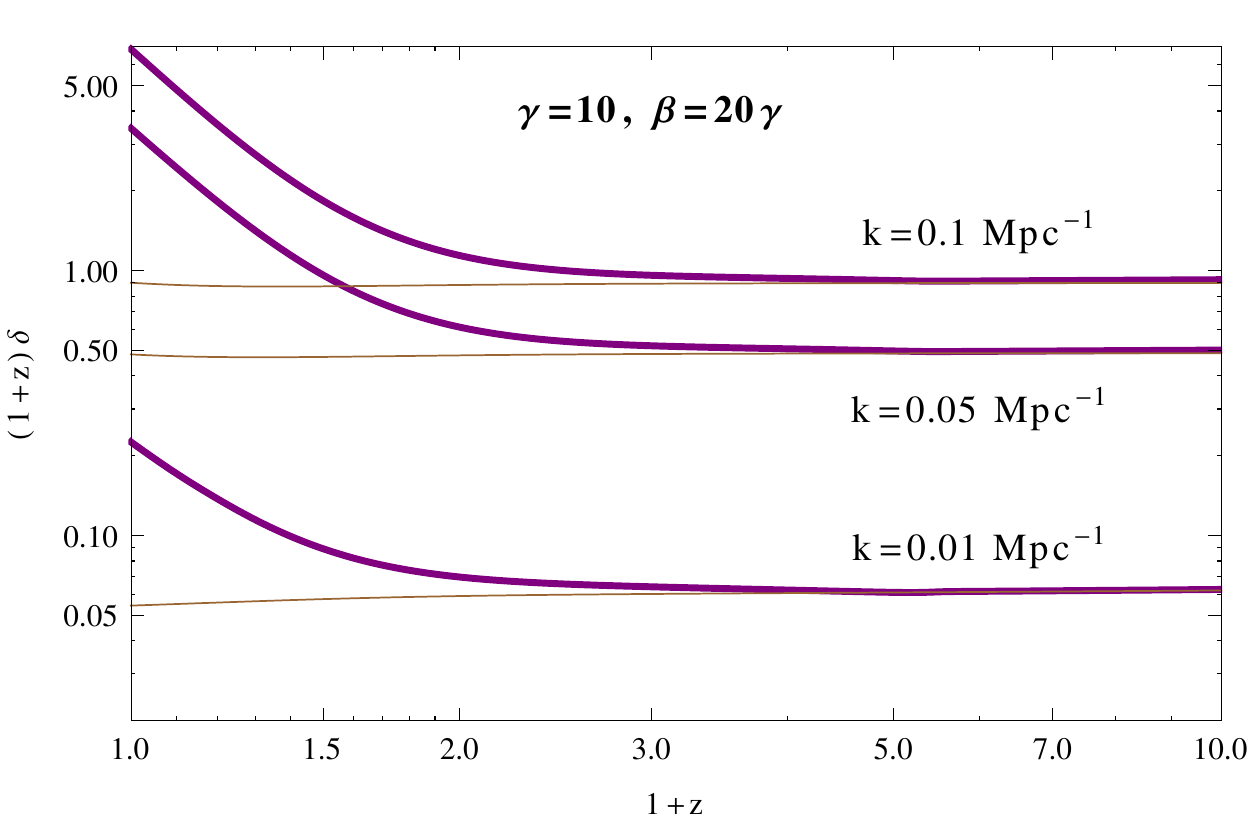}

 \includegraphics[width=\columnwidth]{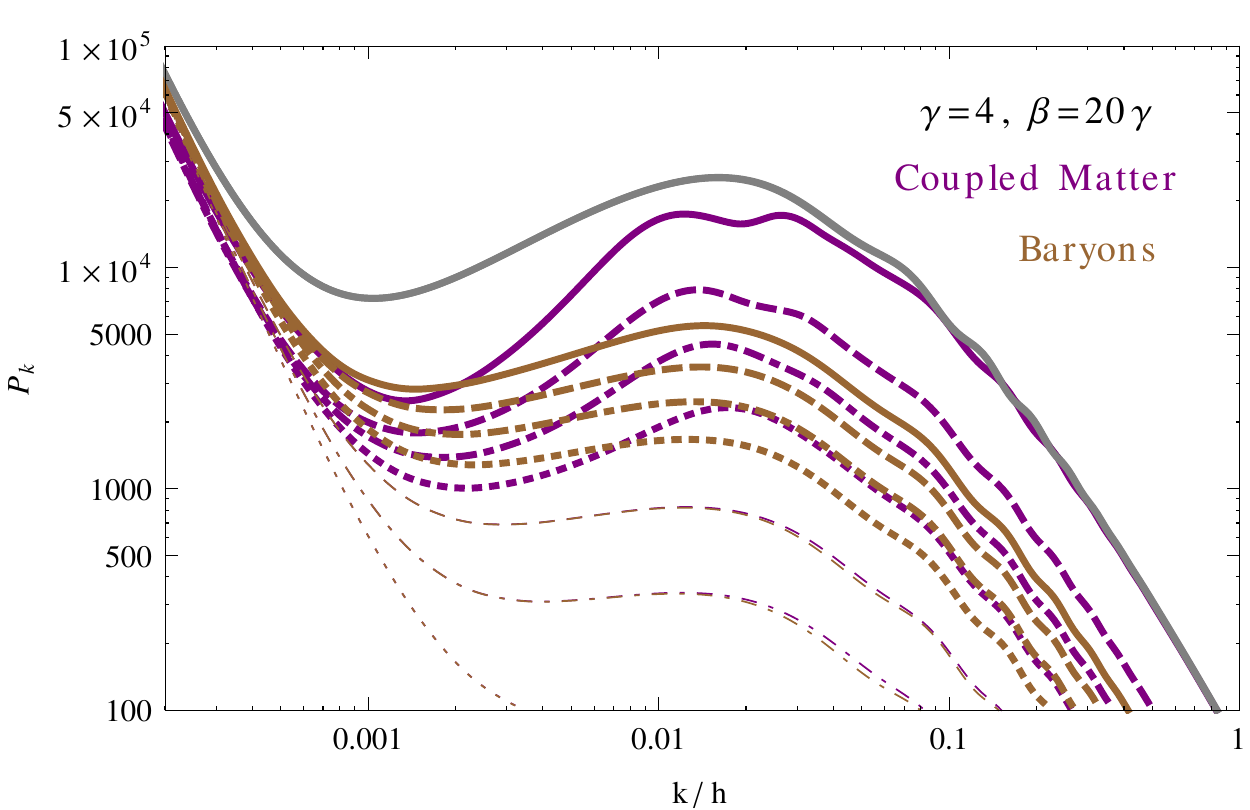}
 \includegraphics[width=\columnwidth]{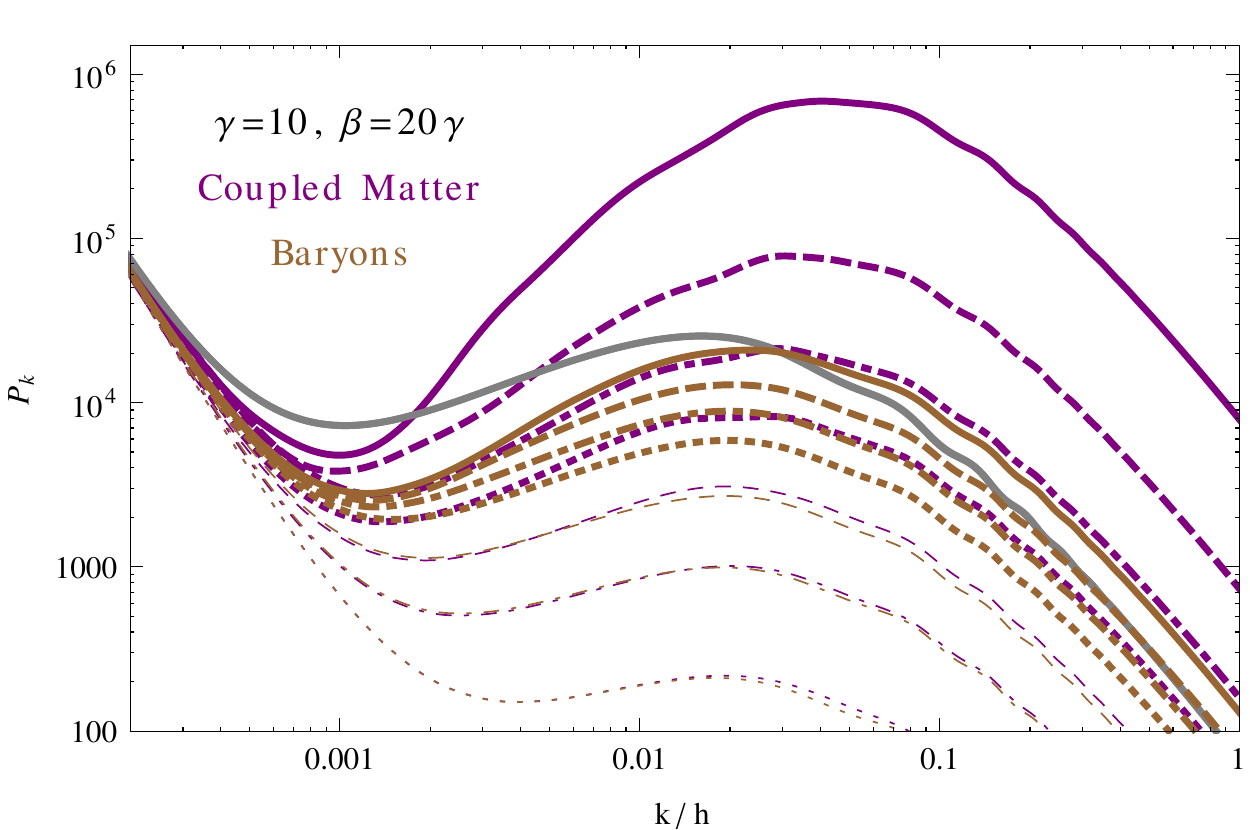}
\end{center}

\caption[Growth of structure for disformally coupled matter]{\justifying
Growth of disformally coupled matter and (uncoupled) baryons. {\bf Upper panel:} Thick purple lines coorespond to the disformally coupled matter and thin brown lines to baryons at the same scales. The pressence of early Dark Energy affects the slope at early times. Afterwards, the Dark Matter perturbation starts growing rapidly, as deduced in the small scale approximation (see Figure \ref{Geff}).
{\bf Lower panel:} Evolution of the power spectrum at $z=0$ (thick, solid), 0.3, 0.6, 1 (thick, dotted), 2, 4 and 10 (thin, dotted). The gray line corresponds to a reference $\La$CDM model. Units of $k$ are Mpc$^{-1}$.
\label{growth}}
\end{figure*}

\begin{figure*}
 \begin{center}
 \includegraphics[width=\columnwidth]{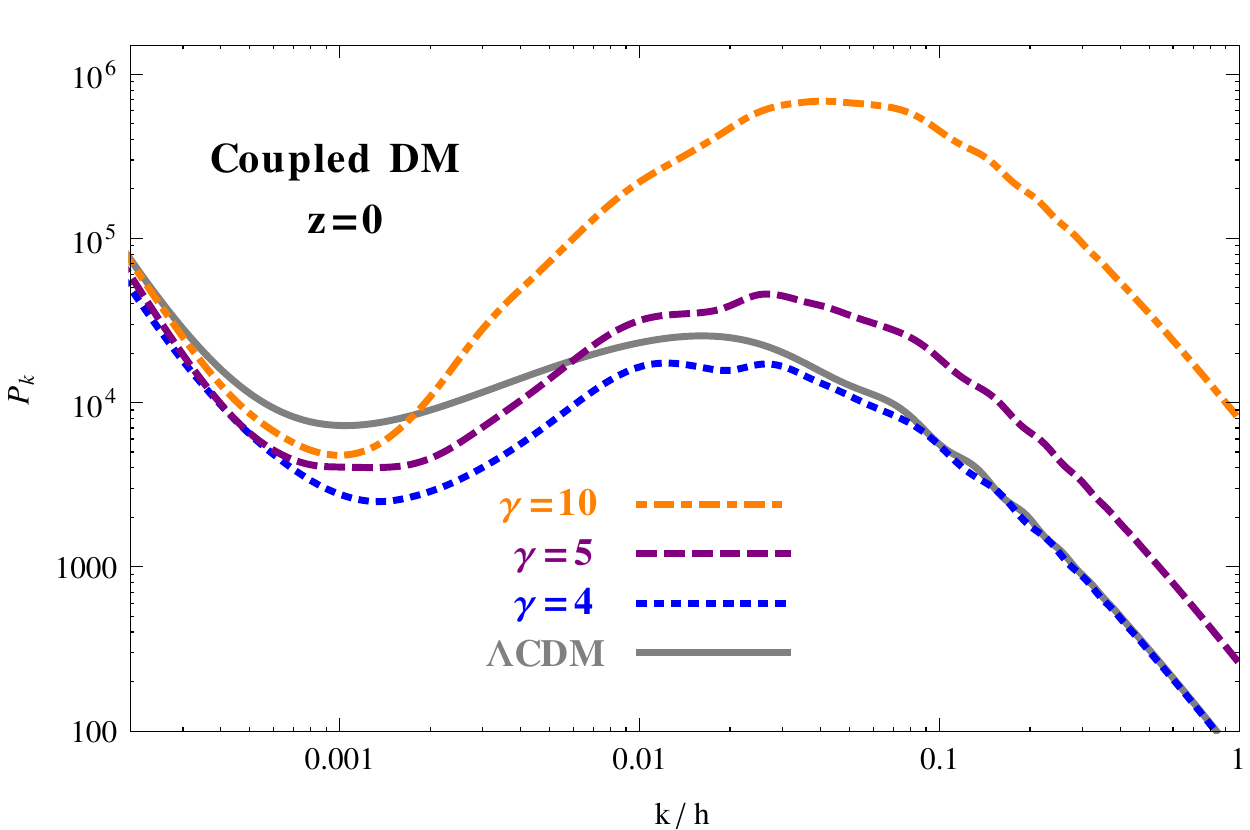}
 \includegraphics[width=\columnwidth]{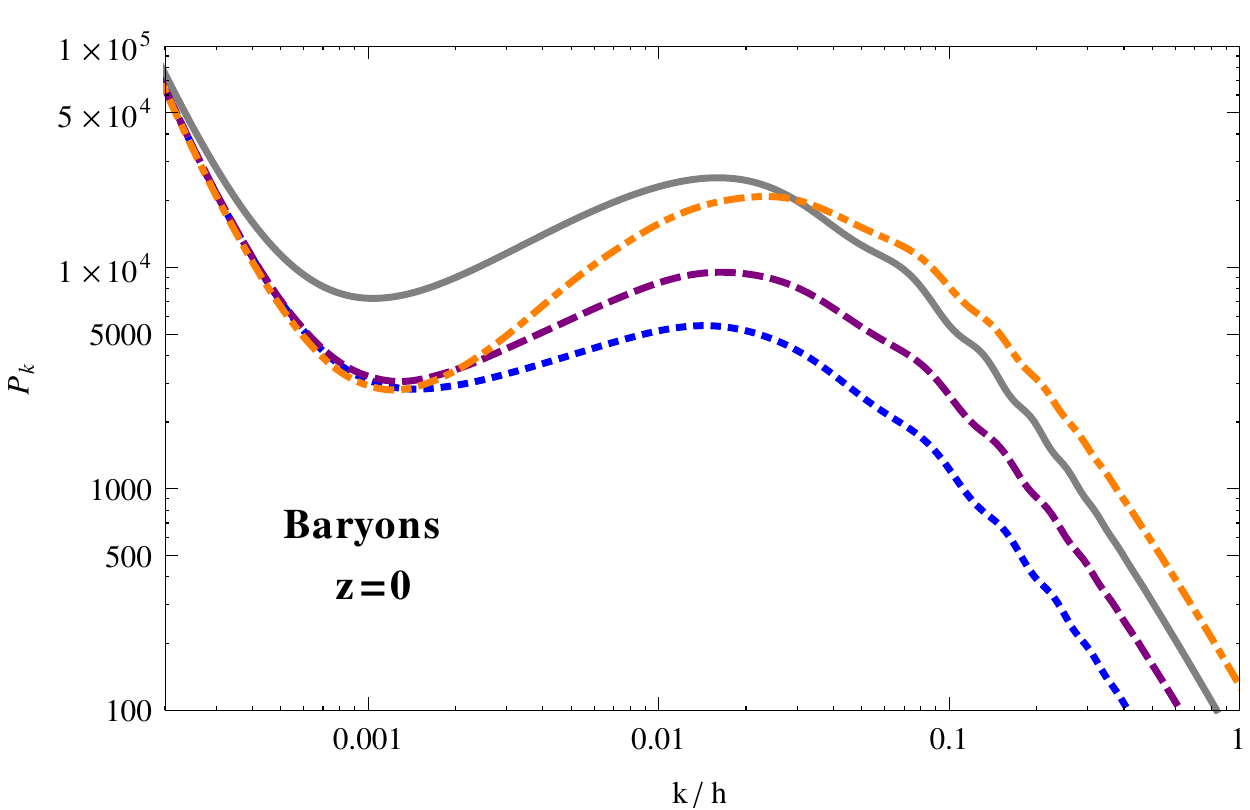}\\
 \includegraphics[width=\columnwidth]{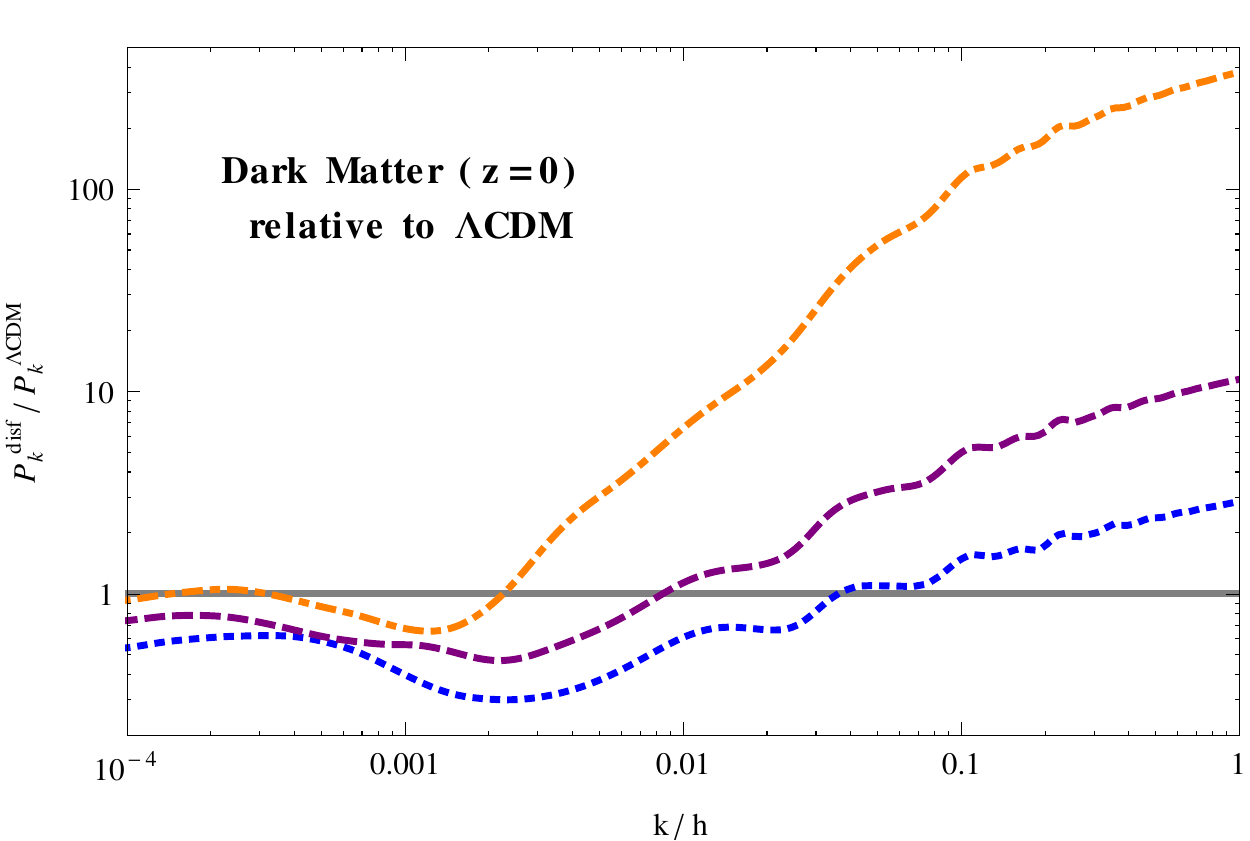}
\includegraphics[width=\columnwidth]{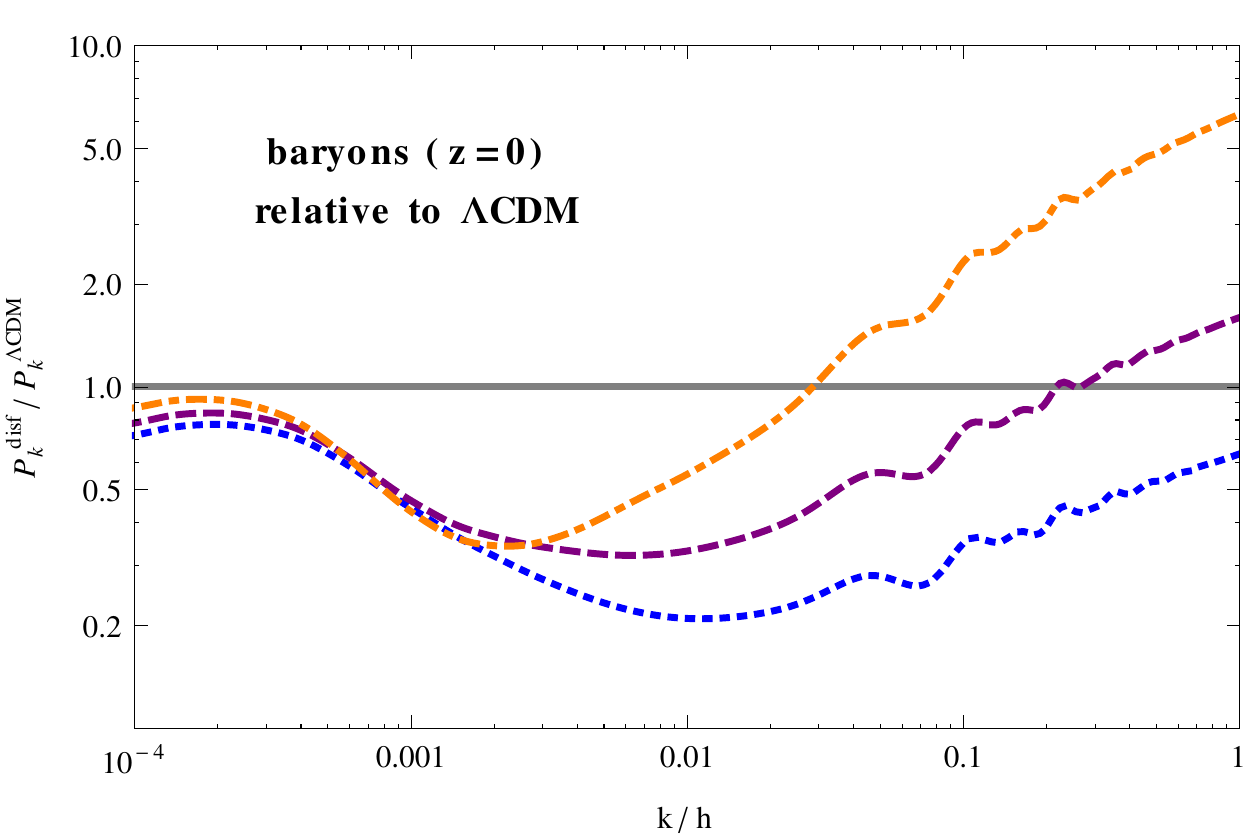}
\caption[Power spectra for disformally coupled matter and (uncoupled) baryons]{\justifying
Power spectra for disformally coupled matter and (uncoupled) baryons at $z=0$ for $\beta= 20\gamma$ and different values of $\gamma$ (units of $k$ are Mpc$^{-1}$). 
{\bf First line:} Power spectrum for (uncoupled) baryons and (coupled) matter.
{\bf Second Line:} Ratio of the power spectra in disformally coupled models relative to $\La$CDM. See the text for a discussion of the different effects.
\label{disf:pk}}
\end{center}
\end{figure*}

\begin{figure*}
 \begin{center}
\includegraphics[width=\columnwidth]{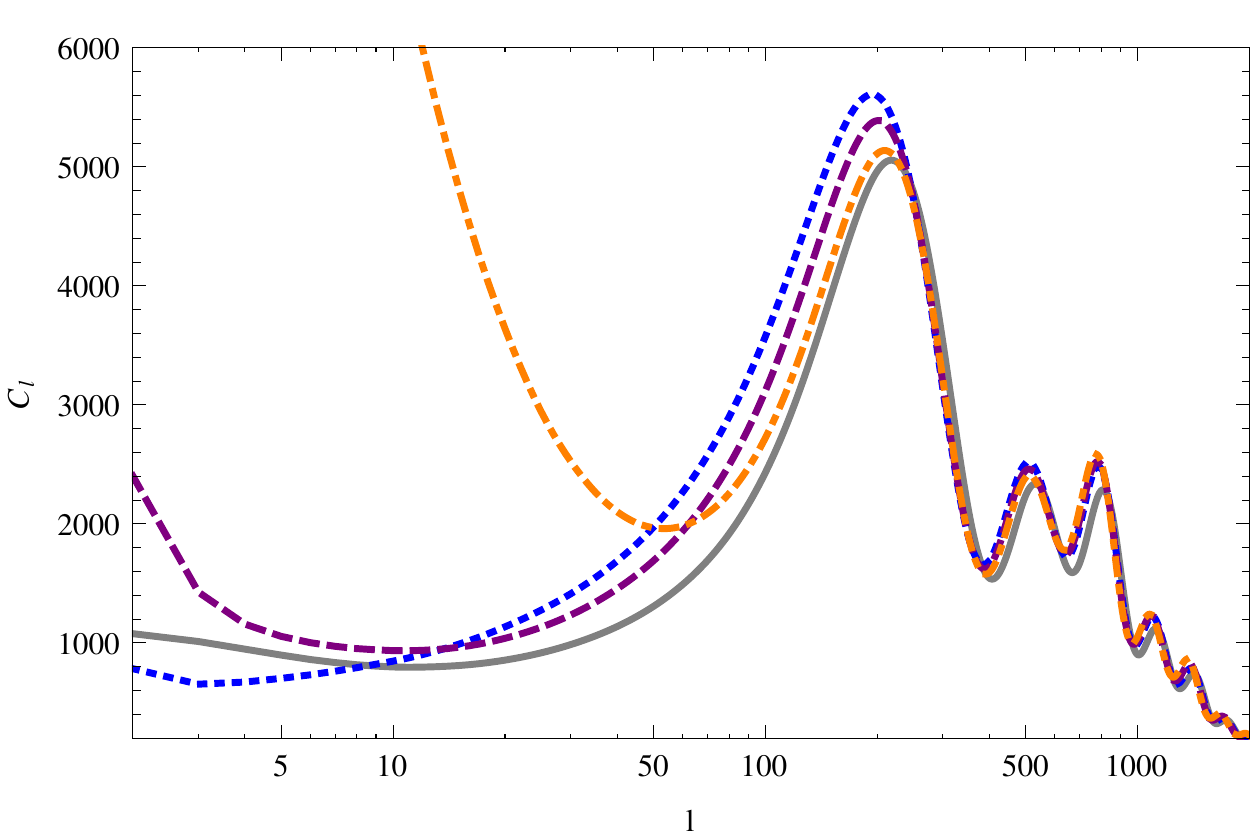}
\includegraphics[width=\columnwidth]{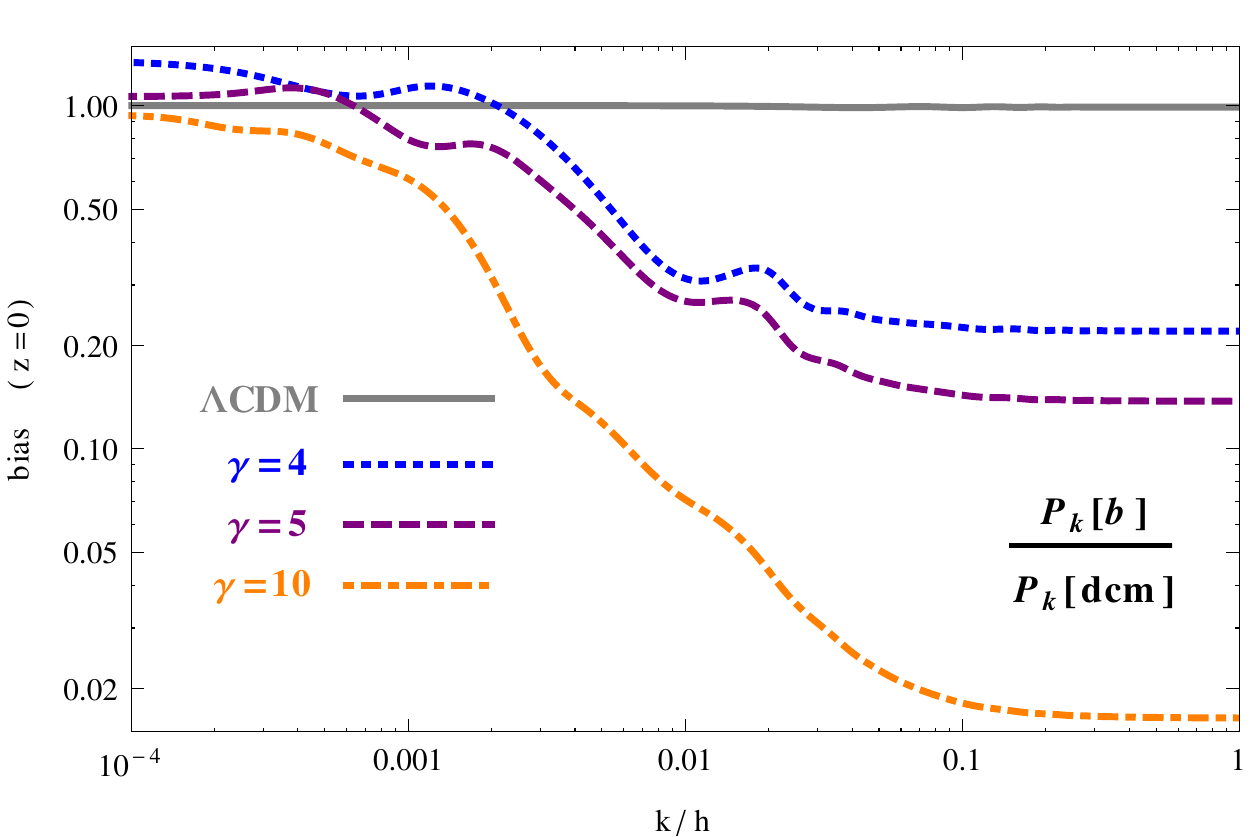}
\caption[CMB power spectrum and baryon-matter bias induced by the coupling]{\justifying
CMB power spectrum (left) and bias between baryons and Disformally Coupled DM (DCM) induced by the coupling (right). The enhanced growth of Dark Matter structures on small scales produces a very large ISW effect. Note that the departures are worse for models with \emph{less} early Dark Energy (higher $\gamma$), as derived in the small scale approximation (\ref{disfcoup:smallscale}). Units of $k$ are Mpc$^{-1}$.
\label{disf:cmb}}
\end{center}
\end{figure*}

The full system of linearized equations (\ref{kgp}-\ref{perturb-disformal}) was solved numerically using a modified version of the Boltzmann code CMBeasy adapted to the Disformally Coupled Dark Matter model described in Section \ref{section:examplemodel}. 
Since matter is essentially uncoupled until $z\lesssim 10$ there was no need to modify the initial conditions, which have been assumed adiabatic.
Figure \ref{growth} shows the evolution of the density contrast of disformally coupled matter. The baryons, which are uncoupled in this particular example, are also shown for comparison. Figure \ref{disf:pk} displays the power spectra for disformally coupled matter and baryons at $z=0$ for different values of the parameters. Figure \ref{disf:cmb} shows the CMB power spectrum and the baryon-DM bias induced by the coupling at $z=0$.

Besides the effect of early Dark Energy and late time scalar force captured in $G_{\rm eff}$, the disformal coupling causes a considerable integrated Sachs-Wolfe effect, a fundamental bias between disformally coupled matter and baryons and large scale oscillatory features beyond the BAO scale. The numerical results and the discussion are restricted to the DCDM model, and focus on the role of the potential slope $\ga$, which mostly determines the late time value of $G_{\rm eff}$. It remains to be studied whether or not similar effects occur in viable models such as the ones described in Section \ref{section:disfviable}, and to what extent they might be observable by current or future surveys.

\setcounter{paragraph}{0}

\paragraph{Early Dark Energy:}
Both the baryons and the coupled Dark Matter are indistinguishable as long as the coupling is negligible. They are equally affected by the presence of early Dark Energy (see Figure \ref{growth}), which produces a departure from the matter domination result $\delta \propto a$: EDE increases the expansion rate without clustering, reducing the formation of structure. This effect was also found for the uncoupled scalar field \cite{Koivisto:2008ak,Zumalacarregui:2010wj}, and is most noticeable for models with higher $\Omega_{\rm ede}$ (e.g. $\gamma=4$).

\paragraph{Late Enhanced Growth:}

The growth of structure is enhanced after the transition takes place, consistently with the small scale approximation (\ref{disfcoup:smallscale}). The large value of $G_{\rm eff}/G$ overcomes the additional friction term, and structures form much faster than in the standard CDM scenario. Models with less EDE suffer a \emph{higher} enhancement, because the effective gravitational constant $G_{\rm eff}\propto\gamma^2$ is larger and the transition occurs earlier (i.e. the field takes longer to dominate the energy content). The effect from the bump in the effective Newton's constant associated to the transition is not obvious in the evolution of $\delta$, and is subdominant with respect to potential domination.

The enhanced growth effect is partly canceled by the early Dark Energy damping. This degeneracy causes the relative resemblance between DCDM power spectra with $\gamma=4$ and the DM power for the standard model on small scales, but fails anywhere else. It would be worth exploring this cancellation in a more systematic way (e.g. Markov chain Monte Carlo exploration of the model parameters), which would in turn require a better understanding of the baryon bias induced by the coupling (see below). However, such an exploration is postponed for future work.

Note that Fourier modes reach the non-linear regime earlier due to the enhanced growth. Upon the failure of linear perturbation theory, the disformal screening mechanism explored in Section \ref{section:disfscreening} might hide these dramatic effects and restore the standard growth, softening the deviations on small scales. Although this seems unlikely to save the example model, it might be necessary to take the effect into account to obtain a fair comparison with observations.%
\footnote{Similar enhanced growth effects have been considered in the context of quintessence conformally coupled to neutrinos, where the necessity of non-linear analysis has been pointed out \cite{Wintergerst:2009fh}.}
 
\paragraph{Scale Dependent Growth and Bias:}

The power spectra show scale dependent evolution, as can be seen in the different power spectra normalized to the corresponding $\La$CDM (second line of Figure \ref{disf:pk}). In the standard model, the linear growth factor is scale independent and does not distinguish baryons from Dark Matter. For disformally coupled Dark Matter, the scale dependent growth follows from the $k$-dependent term in the perturbed coupling (\ref{perturb-disformal}). This feature does not appear in phenomenological coupled models, in which the growth of the coupled matter structures is enhanced, but in a scale independent way (cf. \cite{Clemson:2011an}).

The coupling also modifies the relation between baryonic and Dark Matter structures, since DM couples directly to the field while baryons are only indirectly affected. As baryons are dragged into the potential wells created by the coupled matter, they follow a scale dependent growth pattern, delayed with respect to the dominant matter component.
The resulting bias between the two species is shown in Figure \ref{disf:cmb}. The scale dependence of the bias vanishes both on super-horizon scales ($k/h\lesssim 0.001$ Mpc${-1}$) and the small scales ($k/h\gtrsim 0.1$ Mpc${-1}$), which are well described by the scale-independent $G_{\rm eff}$ (\ref{disfcoup:smallscale}). The intermediate region shows the interplay between the scale dependent growth for the coupled matter and the baryons following these structures.

Since galaxies form out of baryons, this fundamental bias modifies the usual DM-galaxy power relation \cite{Kaiser:1987qv}. Such a correction needs to be taken into account when comparing the observed power spectrum with disformally coupled models. On the other hand, other measurements of the matter distribution such as weak lensing would probe the structures formed by both components, and may be used to break the degeneracy.

\paragraph{CMB: Integrated Sachs-Wolfe Effect:}

The enhancement of the perturbations after the transition causes the very large Integrated Sachs-Wolfe effect appreciated in Figure \ref{disf:cmb}, which becomes most noticeable for the models with higher values of $\gamma$. The model with $\gamma=4$ gives a better fit on the $l\lesssim10$ multipoles, but departs considerably in the range $10\lesssim l \lesssim 200$ due to the effect of early Dark Energy after recombination. The model with less early Dark Energy has the opposite problem: it produces a better fit in the intermediate range $100<l<200$ due to the lower amount of early Dark Energy, but the ISW enhancement explodes at lower multipoles due to the higher value of $G_{\rm eff}$. The different amounts of early Dark Energy have an additional effect on the CMB normalization due to the primary Sachs-Wolfe effect: by reducing the potential wells that redshift the photons, $\Omega_{\rm ede}$ acts increasing the height of the peaks.

\paragraph{Oscillatory Features beyond the BAO Scale:}

Oscillatory features can be appreciated in the coupled matter power spectrum on very large scales. These are likely created as field oscillations on scales near $k\sim H(z)$, which are then transferred to the coupled component, when the coupling is active. They are most noticeable for the models with a large early Dark Energy component, e.g. larger field energy density. Although it constitutes a distinctive feature of the model, the oscillations are not significantly imprinted on the baryonic power spectrum. This, together with the large survey volumes necessary to explore such scales would make it difficult to detect them through LSS surveys.
However, the large scale oscillations would be a characteristic signature in models where the disformal coupling is universal, in which the same effects occur to DM and Baryons.

\subsection{Viable Scenarios}\label{section:disfviable}

The study of cosmological perturbations within the Disformally Coupled Dark Matter model (\ref{examplemodel}-\ref{examplemodel3}) shows very drastic departures in the formation of large scale structure, which seem very difficult to reconcile with observations. 
It would be interesting to obtain a more precise quantification of these discrepancies through an MCMC analysis and explore possible degeneracies (e.g. the growth suppression from early Dark Energy and the enhancement from the high $G_{\rm eff}$). However, it is necessary to address the existence of alternative, viable scenarios already at this stage.

Luckily, the action (\ref{disfcoup:action}) is very general and there is considerable room for improvement through different choices of the functions $A,B$ and $\Lag_\phi$. There are at least two possibilities
\begin{itemize}
\item Introduce a modulation in the disformal factor $B(\phi)\to f(\phi)B(\phi)$, to make $Q_0$ small enough after the field enters the slow roll phase.
This modification can render $\delta G_{\rm eff}$ arbitrarily small, except for a relatively short time around the transition (see Figure \ref{frwbackground}). This type of models would allow us to study the effects imprinted by the transition bump without the problems caused by the high $G_{\rm eff}$ at late times.

\item Constructing the field Lagrangian using a disformal metric, as in the uncoupled model presented in Refs. \cite{Koivisto:2008ak,Zumalacarregui:2010wj}. In this model the transition to slow roll would be partially driven by the scalar field Lagrangian itself, and the effects on matter may be significantly reduced. In the minimal prescription, the matter and field Lagrangian are constructed using the same metric (\ref{metrics}) and no extra parameters are introduced. If this model turned out not to be viable, a different disformal metric for the field and the coupled matter would offer an alternative that is able to interpolate between disformal quintessence and the disformally coupled Dark Matter presented here (e.g. different disformal factors with $B^{(m)}=\epsilon B^{(\phi)}$).
\end{itemize}
Other alternatives could be based on the interplay between the conformal and the disformal part of the coupling.
Viable scenarios might be exploited to alleviate the claimed problems of $\Lambda$CDM with small scale structure formation, such as the tension between Dark Matter simulations and observations with regard to both the density profiles of Dark Matter halos and for the number of predicted substructures inside a given host halo, the baryonic Tully-Fisher relation, the constant galactic surface density or the large scale bulk flows (See reference \cite{Perivolaropoulos:2011hp} for a summary and references therein for further details).

As a final remark, let us note that the enhanced growth rushes the modes into entering the non-linear regime at earlier times, breaking down the perturbative approach followed here. As it was explained in Section \ref{section:disfscreening}, the disformal coupling comes equipped with a screening mechanism, that hides the effects of the additional force on dense environments. Addressing the consequences of the disformal screening in a cosmological context would require considering the non-linear backreaction of the field, which is not properly captured in the approximations considered so far. Chameleon-type theories also show a strengthening of the screening when non-linearities are properly taken into account \cite{Mota:2006fz}.

\section{Discussion} \label{disfcussion}

The disformal relation provides a generalization of the conformal transformation. It has been used to construct theories of modified gravity, notably those which produce non-trivial effects on null geodesics, such as varying speed of light and gravitational alternatives to Dark Matter. It also appears in the description of branes embedded in a higher dimensional bulk space, in which the scalar fields represent the brane position in a certain set of coordinates.
The results of the present work concern the set of theories which can be expressed as General Relativity plus a matter Lagrangian, which is constructed using the disformal metric. This provides a generalization of the old school scalar-tensor theories in the Einstein Frame: Test particles follow geodesics which explicitly involve derivatives of the scalar field, and the energy momentum of the field and coupled matter (computed with respect to the gravitational metric) is not conserved separately.

The existence of additional frames, in which only the conformal or disformal part enter the matter action explicitly, provides novel connections between scalar-tensor theories of gravity.
In particular, it is possible to restore the theory to a Jordan Frame representation by reversing the disformal relation, as was shown in Section \ref{section:confframe}.
By doing so, the transformed Einstein-Hilbert term is shown to be equivalent to a quartic DBI Galileon Lagrangian when expressed in a frame in which the disformal coupling is pushed towards the gravitational sector.  The resulting theory has the correct Horndeski form (\ref{eq:hornyLagrangian}), ensuring the second order nature of the equations and the avoidance of Ostrogradski's ghosts. In particular, it introduces a derivative coupling between the scalar field and gravity, together with higher derivative self interactions. These endow the theory with the Vainshtein screening mechanism, which allows the field to cause effects on cosmological scales while remaining undetectable in the Solar System.

The equivalence between certain higher derivative theories (such as DBI Galileons with conformal or minimal coupling to matter) and disformally coupled theories with an Einstein-Hilbert gravitational sector provides new means to analyze this type of models. Although the equations for disformally coupled theories are rather involved, they are much simpler than higher derivative Horndeski theories in the Jordan Frame and highlight the properties of the different terms. Hence the analysis of disformal theories in the Einstein Frame can be regarded as equivalent to (at least) some scalar-tensor theories featuring the Vainshtein screening mechanism. The kinetic mixing between the coupled degrees of freedom makes it necessary to solve for the second time derivatives of the different components. Although this generally requires some assumptions, e.g. about the metric, a general equation without derivatives of the energy-momentum tensor can be obtained (\ref{cov-field}).
Once solutions are computed, it is possible to restore to the Jordan metric in order to interpret the results and compare to observations.

In high density environments (as measured by the condition $B\rho\gg1$) the field does not feel the presence of disformally coupled, non-relativistic matter.
This provides a novel {\it disformal screening mechanism}, which is distinct from screening mechanisms based on the field potential (Chameleon and Symmetron \cite{Wang:2012kj}).
Our mechanism relies on the existence of a well defined limit $\rho\to\infty$ in the scalar field equation,%
\footnote{One may show that this limit is independent of the assumption of canonical kinetic term for the scalar field we adopted for most of our discussion in this article.} given by equation (\ref{time}), for which the field evolution is independent of the matter distribution and the field gradients (up to effects of order $\sim p/\rho$, $v/c$). If the conformal part $A$ is negligible, only a friction term remains and the field coupling density (\ref{mattereq}) is a decreasing function of time. As it evolves below its cosmological value (provided $V'<0$ and $B'/B>0$), the effects of the coupling are suppressed by a factor $\sim \rho_0/\rho$ and the theory is consistent with precision gravity tests.
Potentially detectable signatures may be obtained in the presence of matter velocity flows, radiation pressure or relativistic matter, strong gravitational fields or gradients of cosmological origin.

The disformal screening mechanism is also related to the Vainhstein effect, which suppresses the gradients of the scalar field and hides the additional force near massive sources due to the higher order derivative self-interactions. The existence of a Vainshtein radius at which the asymptotic solution breaks down can be derived in the Einstein Frame by considering static, vacuum solutions. This property holds if the disformal coupling has the opposite sign than postulated when studying the disformal screening mechanism, and therefore the two effects might be incompatible, at least for the simple models considered here.
In the disformal case, the screening relies on the kinetic mixing between the scalar field and the coupled degrees of freedom, which ultimately allows the existence of a well defined $\rho\to\infty$, non-relativistic limit in which the field is free, or only subject to friction (up to conformal interactions). 
Therefore, both the disformal and Vainshtein mechanisms belong to the kinetic screening category, as they rely on the form in which the field derivatives occur in the action. In both cases, the scale at which the screening takes place is determined by the coefficient of the disformal coupling $B\sim M^{-4}$.

The disformal coupling offers interesting possibilities to build models for cosmic acceleration. In the FRW approximation, the same properties that gave rise to the disformal screening mechanism make the background coupling approximately proportional to the Dark Energy density (\ref{QdFRW}) rather than to the coupled matter energy density. This provides a concrete realization of a class of interacting Dark Matter models which have been extensively studied using phenomenological parameterizations.
The equations for linear perturbations around FRW contain scale dependent terms. These are absent in the pure conformally coupled case, and have hence the potential to distinguish the two possibilities. An analytic equation for the coupled matter perturbations was derived in the small scale limit. On top of an additional friction term, the effect of the fifth force can be encapsulated in the definition of an effective gravitational constant (\ref{disfcoup:smallscale}) which depends on the background coupling factor.

In order to investigate the cosmological implications of a disformal coupling in a simple setting, a \emph{Disformally Coupled Dark Matter} (DCDM) example model was proposed. This has the advantage of avoiding the subtleties of the Einstein Frame description, since gravity, baryons and photons share the same physical metric. 
A DCDM model with exponential functions and no conformal coupling (\ref{examplemodel}-\ref{examplemodel3}) provides a Dark Energy model that tracks the dominant energy component at early times. When the coupling to Dark Matter becomes active, the scalar field enters a slow roll phase in order to dynamically avoid a singularity of the disformal metric. The free parameters can be constrained by observations, and the model is successful at the background level.
When perturbations are included, the DCDM model introduces a series of new effects. The effective gravitational constant for this model is too large, due to the persistence of the coupling at late times and the domination of the scalar field in the energy budget. This causes a too large enhancement of the growth factor, which affects the normalization of the DM and baryon power spectra, producing a very large ISW effect. Scale dependent effects are reflected on matter oscillatory features on very large scales and a scale dependent bias between the coupled Dark Matter and (uncoupled) baryonic component.


There is considerable freedom in the model to produce cosmologically viable scenarios. Models of the DCDM type with less dramatic growth of perturbations can be constructed by modifying the functional dependence of the disformal coupling (e.g. tuning it to become negligible after the transition to slow roll), the scalar field Lagrangian (e.g. constructing it with a disformal metric), or perhaps by the interplay between the conformal and the disformal parts of the coupling. Unlike in the conformal case, the disformal coupling affects ultrarelativistic species and hence variations or extensions of DCDM may postulate or include disformally coupled neutrinos or photons. 

Another phenomenological direction is to consider the disformal screening mechanism in detail.
The results presented here considered a purely disformal coupling, monotonically increasing with the field. Certainly, including a conformal coupling and more general functional forms is of interest. These considerations might help to avoid the gradient instability caused by the Einstein Frame pressure if $Bp>1$, as it was discussed in Section \ref{Bp_instab}.
Once different set-ups are formulated, it is worth to explore the observable signatures for the model by quantifying the effects outlined at the end of Section \ref{fielddecoupl}.

The dependence of the free functions in the Horndeski Lagrangian (\ref{LH4}, \ref{LH5}) on the field kinetic term $X$ has a very special role, as it relates the coupling to gravity to the coefficients of the second derivative field terms. Therefore, it would be worth considering the transformations between frames in the more general case in which the disformal relations are allowed to depend on $X$. The computation of the Ricci scalar associated to this general disformal metric would provide the Jordan Frame representation of the most general scalar-tensor theory that accepts an Einstein Frame description. Since the equations simplify considerably in this frame, the phenomenology of these theories would be relatively easy to address.

Finally, the existence of a well behaved $\rho\to \infty$ classical limit in the field equation suggests that disformally coupled theories might introduce new interesting features for the physics of gravitational singularities and other high energy regimes. The implications of kinetic mixing for the formation of black holes or the origin of the universe is beyond the scope of the present work, but it might provide a fruitful exploration to pursue in the future.
This discussion provided just a glimpse to the potential applications of the disformal relation. As a generalization of the conformal case, which was very central to the development of gravitation and cosmology in the $20^{\rm th}$ Century, the use of disformal transformation might provide novel ways to address the gravitational physics of the $21^{\rm th}$ Century.

\acknowledgments{We thank L. Amendola, A. Barreira, J. Beltran-Jimenez, J.A.R. Cembranos, F. de Juan, A.L. Maroto, J. Sakstein and I. Sawicki for discussions, J. Garc\'ia-Bellido for a careful reading of the manuscript and useful comments, R. Gannouji and S. Renaux-Petel for correspondence and comments to the first version and the anonymous referee for very insightful comments. TK and DFM are supported by the Norwegian research council. MZ is supported by the Madrid Regional Government (CAM) under the program HEPHACOS S2009/ESP-1473-02 and enjoyed a Yggdrasil mobility grant at the University of Oslo during the first stages of this project.}

\appendix

\section{Disformal Relations} \label{disfappendix}


Consider the disformal relation between two metrics, specified by the two scalar functions $A$, $B$, and a vector $b_\mu$,
\begin{equation}
\bar{g}_{\mu\nu} = A g_{\mu\nu}+ B b_\mu b_\nu\,.
\end{equation}
The inverse metric can be found by contraction 
\begin{equation}\label{inverseapp}
\bar{g}^{\mu\nu} = \frac{1}{A}\lp g^{\mu\nu}- \ga^2 b^\mu b^\nu \rp\,.
\end{equation}
where 
\begin{equation}
\gamma^2 \equiv \frac{B}{A+B b^2}\,,
\end{equation}
and $ b^2 \equiv g^{\mu\nu}b_\mu b_\nu \equiv b^\mu b_\mu$.
The determinant of the barred and unbarred metrics are related
\begin{equation}\label{determinant2}
\sqrt{\frac{\bar{g}}{g}}=A\sqrt{\frac{AB}{\gamma^2}} = A^2\sqrt{1+\frac{B}{A}b^2} \,,
\end{equation} 
The above relation is derived in Appendix C of Ref. \cite{Bekenstein:2004ne}.

It is possible to write the relation of stress energy momentum tensor (associated to a Lagrangian $\sqrt{-g}\Lag$) in the two metrics by using the chain rule
\begin{equation}
 T^{\mu\nu} \equiv \frac{2}{\sqrt{-g}}\frac{\delta\lp\sqrt{-g}\mathcal{L}\rp}{\delta g_{\mu\nu}} 
=\sqrt{\frac{\bar g}{ g}}  \frac{\delta \bar g_{\al\bt}}{\delta g_{\mu\nu}} 
\lp\frac{2}{\sqrt{-\bar g}}\frac{\delta\lp\sqrt{-g}\mathcal{L}\rp}{\delta \bar g_{\al\bt}}\rp \,.
\end{equation}
By identifying the quantity in brackets as $\bar T^{\mu\nu}$ and using (\ref{determinant2}), the following relation follows
\begin{equation}
T^{\mu\nu} = A^3\sqrt{1+\frac{B}{A}b^2} \,\bar{T}^{\mu\nu}\,.
\end{equation}
The equivalent relation with lower indices is considerably more involved
\begin{equation}\label{tensortransform}
T_{\mu\nu} = \sqrt{\frac{\bar{g}}{g}}D_{\mu\nu}^{\phantom{\mu\nu}\alpha\beta}\bar{T}_{\alpha\beta}\,,
\end{equation}
where
\begin{equation}
 D_{\mu\nu}^{\phantom{\mu\nu}\alpha\beta}
\equiv \frac{\delta \bar{g}^{\alpha\beta}}{\delta g^{\mu\nu}}
= \frac{1}{A}\lp\delta^\alpha_\mu\delta^\beta_\nu- 2\ga^2 b^\alpha b_{(\mu}\delta^\beta_{\nu)}+ \ga^4 b_\mu b_\nu b^\alpha b^\beta\rp\,.
\end{equation}

The inverse relations are provided below for completeness
\begin{equation}
 g_{\mu\nu}  =  \frac{1}{A}\lp\bar{g}_{\mu\nu} -B \bar{b}_\mu \bar{b}_\nu\rp \,, 
\end{equation}
\begin{equation}
g^{\mu\nu}  =  A\lp \bar{g}^{\mu\nu}+\bar{\ga}^2\bar{b}^\mu\bar{b}^\nu\rp\,, \quad \bar{\ga}^2 \equiv \frac{B}{A-B\bar{b}^2}\,,
\end{equation}
\begin{equation}
\bar{D}^{\mu\nu}_{\phantom{\mu\nu}\alpha\beta}  =  A\lp \delta^\mu_\alpha\delta^\nu_\beta + 2\bar{\ga}^2\delta^{\mu}_{(\alpha} \bar{b}_{\beta)}\bar{b}^\nu
+ \bar{\ga}^4 \bar{b}^\mu\bar{b}^\nu \bar{b}_\alpha\bar{b}_\beta \rp\,,
\end{equation}
where $\bar{b}^\mu \equiv \bar{g}^{\mu\nu}b_\nu$.
Note that $\bar{b}_\mu=b_\mu$, $\bar{b}^\mu = B/(A\bar{\gamma}^2)b^\mu$ and $\ga^2 b^2=B\bar{b}^2$.

\subsection{Disformal Geodesics} \label{app:geodesics}

The expression for the disformal connection (\ref{connection}) can be expanded in terms of the functions in the disformal metric
\begin{widetext}
\begin{eqnarray}\label{connectionlong}
\bar{\Gamma}^\mu_{\al\bt} 
& = & \Ga^\mu_{\alpha\beta} + \delta^\mu_{(\al}{\log{A}}_{,\bt)}-\frac{1}{2}{\log{A}}^{,\mu}g_{\al\bt} 
 +  \frac{1}{A}\lp \phi^{,\mu} B_{,(\alpha}\phi_{,\bt)} - \frac{1}{2}B^{,\mu}\phi_{,\al} \phi_{,\bt} \rp  \\
&& - \frac{\ga^2}{A}\phi^{,\mu}\Big[ A_{,(\al}\phi_{,\bt)} - \frac{1}{2}\phi^{,\la} A_{,\la}g_{\al\bt} 
 - 2X\lp B_{,\al}\phi_{,\bt}-\frac{1}{2}\phi^{,\la} B_{,\la}\phi_{,\al} \phi_{,\bt} \rp \Big] \nonumber \\
&& + \frac{B}{A}\Big[ \nabla_{(\al}\lp \phi_{,\bt)}\phi^{,\mu}\rp - \frac{1}{2}\nabla^\mu\lp \phi_{,\al} \phi_{,\bt}\rp - \ga^2 \phi^{,\mu} \phi^{,\la} \lp \nabla_{(\al}\lp \phi_{,\bt)}\phi_{,\la}\rp - \frac{1}{2}\nabla_\la\lp \phi_{,\al}\phi_{,\bt} \rp \rp\Big]\,. \nonumber
\end{eqnarray}
Here $\ga^2 \equiv \frac{B}{A-2BX}$ arises from the inverse barred metric, eq. (\ref{inverseapp}).
The first term is just the connection of the unbarred metric, and the two following terms arise from the purely conformal transformation involving derivatives of $A$. The fourth term and the second line contain the first order derivative terms from the disformal contribution to the metric $B$. The third line shows the second order derivative terms $\nabla\nabla \phi$.

\section{General Perturbations} \label{generalDelQ}

The coupling density perturbation that enters the linear equations (\ref{kgp}, \ref{cp}) in the case where both the conformal and the disformal parts of the coupling are relevant has the rather complicated form
\begin{equation}
\delta Q = Q_{\rho} \delta_{\rm dc} + Q_{\phi} \delta\phi +  Q_{d\phi}\delta\dot\phi  + Q_{\Phi} \Phi + Q_{\dot \Psi} \dot \Psi \,,
\end{equation}
where
\begin{eqnarray}
 Q_{\rho} &=& \rho \left(1-\frac{B}{A} \dot \phi ^2 \right)\frac{ A' (A-2 B \dot \phi^2) +B' \dot \phi^2 - A (2 B V'+ 6 B H \dot \phi )}
{2 \left(A+B \rho -B\dot\phi^2\right)^2} \,, 
\end{eqnarray}\begin{eqnarray}
 Q_{\phi} &=& 
 \Big( (1-2 \frac{B}{A} \dot\phi^2)A'' + B'' \dot\phi^2
- 2 B \left(\frac{k^2}{a^2}+V''\right) \Big) \frac{\rho}{ 2A (A+B \rho -B \dot\phi^2)}  
\nonumber \\ & &
- \Big(\lp\frac{A'}{A}\rp^2 (A^2-2 B (2 A+B \rho ) \dot\phi^2+2 B^2 \dot\phi^4)
+ A' \left(-2 B (V'+3 H \dot\phi)+B'(\rho +2 \dot\phi^2)\right)
\nonumber \\ & &
+ 2 A B' (V'+3 H \dot\phi)+A^2 B'^2 \dot\phi^2 (\rho -\dot\phi^2)\Big) \frac{\rho}{2 (A+B \rho -B \dot\phi^2)^2}
\,, 
\end{eqnarray}\begin{eqnarray}
 Q_{d\phi} &=&  -\frac{ B (A+2 B \rho ) \frac{A'}{A} \dot\phi -(A+B \rho ) B' \dot\phi+B \left(2 B V' \dot\phi+3 H(A+B \rho +B \dot\phi^2)\right)}{ \left(A+B \rho -B \dot\phi^2\right)^2} \rho
\,, 
\end{eqnarray}\begin{eqnarray}
 Q_{\Phi} &=&  \frac{B (A+2 B \rho ) \frac{A'}{A} \dot\phi 
-(A+B \rho ) B' \dot\phi+2 B \left(3 H (A+B \rho )+B V' \dot\phi\right)}
{(A+B \rho -B \dot\phi^2)^2}\rho  \dot\phi \,, 
\end{eqnarray}\begin{eqnarray}
 Q_{\dot\Psi}  &=&  \frac{3 B \rho  \dot\phi}{A+B \rho -B \dot\phi^2} \,.
\end{eqnarray}
The equations for the perturbations can also be found in Ref. \cite{vandeBruck:2012vq} for the case in which the coupled fluid is allowed to have pressure.

\section{Dynamical system analysis} \label{dynamicalsystem}

It is useful to reformulate the system in terms of the dimensionless variables
\be
\Omega=\frac{8\pi G\rho}{3H^2}\,, \quad X=\sqrt{\frac{4\pi G}{3}}\frac{d\phi}{dN}\,, \quad Y = \frac{8\pi G V}{3H^2}\,, \quad
Z=\frac{BH^2}{8 \pi G}\,,
\ee
and using the e-folding time $N=\log{a}$ as the time variable. 
The Friedmann constraint then reads
\be \label{fcon}
1=\Omega+X^2+Y\,.
\ee
For concreteness, we assume the exponential forms
\be 
A=A_0 e^{\alpha\phi/M_p}\,, \qquad
B=B_0 e^{\beta\phi/M_p}\,,  \qquad
V=V_0 e^{-\gamma\phi/M_p} \,,
\ee

with $M_p=\lp4 \pi G/3\rp^{-1/2}$. {Note that when $\al=\bt$ one has a simpler form 
$$ \bar{g}_{\mu\nu} =e^{\sqrt{\frac{4}{3} \pi G}\alpha\phi}\lp g_{\mu\nu} + \frac{1}{M^4}\phi_{,\mu} \phi_{,\nu}\rp \,. $$}

We use Eq. (\ref{fcon}) to eliminate $\Omega$ from the system. Equations of motion can then be rewritten in terms of the remaining variables as
\ba
A' & = & \al A X\,, \label{eeA} \\
X' & = & 
\frac{1}{4}\Big[A-3 Z \left(3 X^2+Y-1\right)\Big]^{-1} \Big[ A \left(6 X^3+X^2 \alpha -6 X (Y+1)+(Y-1) \alpha -2 Y \gamma \right)
\nonumber \\ 
& - & 6 X Z \left(9 X^4+X^3 (2 \alpha -\beta
   )-6 X^2 (Y+1)+X (Y-1) (2 \alpha -\beta )-2 X Y \gamma -3 (Y-1)^2\right)\Big]
\nonumber \\
Y' & = & 3\left(1+ X^2+ \frac{1}{3}\gamma X- Y\right)Y\,, \\
Z' & = & -3\left(1+X^2-\frac{1}{3} \beta X -  Y \right)Z  \,. 
\ea
The fixed points are:
\begin{itemize}

\item Matter domination: $\Omega=1$, $X=0$. This solution is always a saddle point, since the eigenvalues corresponding to it are $(3,-3,-\alpha/2)$.

\item Scaling solution: $\Omega= 1 - \al^2/4$, $X = -\al/2$. This point in the phase space is never an attractor.

\item Conformal scaling solution: $\Omega=2 \lp\frac{36 + 3 \al \ga - 2 \ga^2}{3 \al - 2 \ga}\rp^2$, $X=\frac{6}{3 \al - 2 \ga}$. The eigenvalues are
$$(\frac{6 (\beta +\gamma )}{\alpha -2 \gamma },\frac{\pm\sqrt{(\alpha -2 \gamma )^2 \left(-\left(4 \alpha ^2+63\right) \gamma ^2+\alpha  \left(\alpha ^2-54\right) \gamma +45
   \alpha ^2+4 \alpha  \gamma ^3+1296\right)}-3 (\alpha -2 \gamma ) (\alpha -\gamma )}{2 (\alpha -2 \gamma )^2}
)\,.$$
The general form of the stability condition is too messy to write down here.

\item Kinetic domination: $\Omega=0$, $X=\pm 1$. This solution is stable given $\pm\alpha <-6\land \pm\beta <6\land \pm\gamma <-6$.

\item Scalar dominated solution: $\Omega=0$, $X=-\ga/6$. The stability conditions for this fixed point are modified in the presence of 
the disformal coupling. To be explicit, this solution exists and is stable if either
$\alpha +\frac{36}{\gamma }>2 \gamma \land \gamma <6\land ((\beta >0\land \gamma >0)\lor (-6<\beta \leq 0\land \beta +\gamma
   >0))$ or  $\alpha +\frac{36}{\gamma }<2 \gamma \land \gamma +6>0\land ((\gamma <0\land \beta \leq 0)\lor (\beta
   >0\land \beta +\gamma <0\land \beta \leq 6))$. 

\item Disformal scaling solution: $\Omega=(36 - \bt^2 \pm \bt \sqrt{\bt^2-36})/18$, $X=(\bt \pm \sqrt{\bt^2-36})/6$. This fixed point exists when $\beta>6$, but it is never stable. 

\end{itemize}
\end{widetext}
The last one of these is new and it exists when $A=1$. The plus-branch is physical, given $\bt>6$.

\bibliographystyle{ieeetr}
\bibliography{aPRDdisf2}

\end{document}